\documentclass[11pt]{article}
\usepackage{ifpdf}
\ifpdf
\usepackage{graphicx,color}
\usepackage{hyperref} 
\else
\usepackage[dvipdfmx]{graphicx,color}
\usepackage[dvipdfmx]{hyperref} 
\fi
\usepackage{amssymb,amsfonts}
\usepackage{amsmath}
\usepackage[capitalise]{cleveref}
\usepackage{epstopdf}

\usepackage{cite}

\makeatletter
\renewcommand{\theequation}{\arabic{section}.\arabic{equation}}
\@addtoreset{equation}{section}
\makeatother
\newcommand{\diff}[3]{
\if 1#1  \frac{{\rm d} #2 }{{\rm d} #3 }
\else  \frac{{\rm d}^{#1} #2 }{{\rm d}#3^{#1} } \fi
}
\newcommand{\tr}[0]{{\rm tr}}
\newcommand{\vph}[0]{\varphi}
\newcommand{\dps}[0]{\displaystyle}
\newcommand{\lam}{\lambda}
\newcommand{\ep}{\epsilon}
\newcommand{\del}{\delta}
\newcommand{\ovl}{\overline}
\newcommand{\si}{\sigma}
\newcommand{\Si}{\Sigma}
\newcommand{\ola}[1]{\overleftarrow{#1}}
\newcommand{\coD}{{\cal D}}
\newcommand{\eqs}[1]{\begin{equation}\begin{split} #1 \end{split}\end{equation}}

\begin{document}
\begin{titlepage}

\begin{flushright}
IPMU15-0035 \\
\end{flushright}

\vskip 1.35cm
\begin{center}

{\large 
{\bf Threshold Corrections to Baryon Number Violating Operators \\ in Supersymmetric $SU(5)$ GUTs}
}

\vskip 1.2cm

Junji Hisano$^{a,b}$, 
Takumi Kuwahara$^a$,
and
Yuji Omura$^{a}$\\

\vskip 0.4cm

{\it $^a$Department of Physics,
Nagoya University, Nagoya 464-8602, Japan}\\
{\it $^b$Kavli Institute for the Physics and Mathematics of the Universe
 (Kavli IPMU),
University of Tokyo, Kashiwa 277-8568, Japan}
\date{\today}

\vskip 1.5cm

\begin{abstract} 

The nucleon decay is a significant phenomenon to verify grand unified theories (GUTs). For the precise prediction of the nucleon lifetime induced by the gauge bosons associated with the unified gauge group, it is important to include the renormalization effects on the Wilson coefficients of the dimension-six baryon number violating operators. In this study, we have derived the threshold corrections to these coefficients at the one-loop level in the minimal supersymmetric $SU(5)$ GUT and the extended one with additional $SU(5)$ vector-like pairs. As a result, it is found that the nucleon decay rate is enhanced about 5\% in the minimal setup, and then the enhancement could become smaller in the vector-like matter extensions.

\end{abstract}

\end{center}
\end{titlepage}

\section{Introduction}
The supersymmetric grand unified theories (SUSY GUTs) are attractive extensions of the Standard Model (SM).
The three gauge groups of the SM are unified into one, and the SM fermions are embedded into the fields charged
under the unified gauge group in the GUT. The minimal candidate for the gauge symmetry is $SU(5)$,
and we may understand the origin of the hypercharge assignment according to the group structure of $SU(5)$.
SUSY also plays a crucial role in the gauge coupling unification as well as
the natural explanation of the gauge hierarchy problem, and we are looking forward to the discovery of the SUSY particles
at the LHC experiment.
In 2012, it was reported that a scalar particle, which may be consistent with the SM Higgs boson, was discovered around 126~GeV \cite{Aad:2012tfa,Chatrchyan:2012ufa}. The SM is firmly established and we expect that
new physics predicted by the SUSY GUT is also discovered near future, although it has not been found yet at the LHC \cite{Chatrchyan:2013xna,Aad:2014kra,Aad:2014vma,Aad:2014wea,CMS:2014dpa,Khachatryan:2014mma,Khachatryan:2014qwa}. 

On the other hand, it is true that there are several issues which should be carefully studied in the SUSY GUTs.
One of  the issues is how to achieve the $126$-GeV scalar boson. The low-energy effective field theory (EFT) for the SUSY GUT is considered to be the minimal supersymmetric standard model (MSSM). It is known that the MSSM predicts the upper bound on  
the Higgs mass, and the observed Higgs mass may require high-scale SUSY \cite{Giudice:2011cg,Ibe:2011aa,Ibe:2012hu}, or very specific SUSY mass spectrums \cite{Hall:2011aa}, unless the MSSM is further extended, for instance, introducing extra vector-like fields \cite{Martin:2009bg}.

Another big issue is from the experimental constraints on baryon number violation, such as nucleon decay. 
The GUTs unify quarks and leptons, so that the baryon-number-violating processes are introduced through the 
gauge interaction. The processes are strongly suppressed by the GUT scale, but it is possible 
to test the models through the nucleon decay search.
The current status of the nucleon decay experiments is as follows: the partial lifetime limit on $p \to \pi^0 e^+$ is $\tau(p \to \pi^0 e^+) > 1.4 \times 10^{34}$ years \cite{Shiozawa:2013pre,Babu:2013jba}, and the partial lifetime limit on $p \to K^+ \ovl\nu$ is $\tau(p \to K^+\ovl\nu) > 5.9 \times 10^{33}$ years \cite{Abe:2014mwa}.
The prediction of the GUT depends on the scenario between the electroweak (EW) and the GUT scale ($\sim 10^{16}$~GeV). In the minimal SUSY $SU(5)$ GUT, the color-triplet Higgs exchange induces dangerous dimension-five operators to cause baryon number violation  \cite{Sakai:1981pk,Weinberg:1981wj}. It is a serious problem, if the SUSY scale is close to the EW scale. If the SUSY scale is much higher, the constraint from the color-triplet Higgs becomes mild and the dominant decay mode $p \to K^+ \ovl\nu$ may be detected at the future detectors \cite{Hisano:2013exa,Nagata:2013sba}. 
Furthermore, the heavy gaugino masses make the GUT scale lower, so that 
the decay rate for $p \to \pi^0 e^+$, induced by a massive gauge boson ($X$ boson), may be also large enough to be detected at the future detectors \cite{Hisano:2013cqa}. 
If we introduce additional SM-charged fields, the gauge coupling constants would become larger at the GUT scale since the extra fields contribute to the running of the gauge coupling constants \cite{Martin:2009bg}. Then the nucleon decay through the $X$-boson exchange is enhanced \cite{Hisano:2012wq}. 
Note that the lifetime of proton is very sensitive to the $X$-boson mass, because the decay width is suppressed by the fourth power of the $X$-boson mass. This means that we need careful analysis to draw the constraint on the $X$ boson.

In this paper we derive the threshold corrections to the Wilson coefficients of the baryon-number violating dimension-six operators induced by the $X$ boson in the minimal setup of the $SU(5)$ GUT and the extended one with extra $SU(5)$ vector-like pairs. 
In particular, since the unified gauge coupling at the GUT scale becomes large in the vector-like extensions,
it is important to evaluate quantum corrections via gauge interaction in these models.
The two-loop order corrections to the dimension-six operators have been investigated, including the long-distance effect \cite{Nihei:1994tx} and the short-distance effect \cite{Hisano:2013ege}. However, the threshold corrections to the dimension-six operators at the GUT scale have never been discussed. The correction will not be non-negligible, especially  when the gauge coupling constants at the GUT scale are large. We evaluate the corrections at the one-loop level analytically.

This paper is organized as follows: in \cref{sec:GUTs}, we introduce the minimal SUSY $SU(5)$ GUT to summarize our notations. 
In \cref{sec:radiative}, we show the radiative corrections such as the wave function renormalizations, vertex corrections, and box-like corrections, using supergraph techniques. The definition of covariant derivatives on superfields in this paper is the same as in Ref.~\cite{Martin:1997ns} though we use the metric signature $\eta_{\mu\nu}={\rm diag}(1,-1,-1,-1)$. We adopt the $\ovl{\text{DR}}$  scheme \cite{Siegel:1979wq} for the gauge coupling constants while we impose the on-shell condition to the $X$ boson mass $M_X$. For simplicity, we choose the Feynman gauge ($\xi=1$) through this paper. In the next section, we estimate the threshold corrections to the Wilson coefficients of the dimension-six operators at the GUT scale, and we evaluate the numerical results for these finite corrections in the minimal SUSY $SU(5)$ GUT and its vector-like matter extensions. Finally, we summarize our paper in \cref{sec:conclusion}.
We introduce the gauge interactions relevant to our analysis in \cref{ap:gauge int}.
Our explicit results on the one-loop corrections are shown in \cref{app:explicit_loops}, and
the renormalization group equations (RGEs) of gauge couplings, Yukawa couplings and the Wilson coefficients for dimension-six operators are discussed in \cref{app:RGE}.

\section{SUSY $SU(5)$ GUTs \label{sec:GUTs}}
In the SUSY extensions of the SM, it is useful to use the superfield formalism in order to describe the fundamental interactions. Matter fields, Higgs fields, and their superpartners are embedded in chiral superfields and their conjugation. Gauge bosons and gauginos are described by vector superfields. 

In the SUSY extension \cite{Sakai:1981gr} of the minimal $SU(5)$ GUT \cite{Georgi:1974sy}, the matter fields are given by the $\ovl{\bold 5}$ and $\bold{10}$  representational  superfields which are denoted by $\Phi$ and $\Psi$ as follows:
\eqs{
\Phi_{iA}(\bold{\bar 5})=\left(
\begin{array}{c}
D^C_{i\alpha} \\
\ep_{rs} L^s_i
\end{array}\right), ~~~~~ 
\Psi^{AB}_i(\bold{10})= \frac{1}{\sqrt 2}\left(
\begin{array}{cc}
\ep^{\alpha\beta\gamma} e^{-i\vph_i} U^C_{i\gamma} & Q^{r\alpha}_i \\
- Q^{s\beta}_i & \ep^{sr} V_{ij} E^C_j
\end{array}
\right),
}
where $A, B, \cdots = 1, 2, \cdots, 5$ are the indices of the $SU(5)$, $\alpha, \beta, \cdots = 1, 2, 3$ and $r, s, \cdots = 1,2$ are the indices of the $SU(3)_C$ and $SU(2)_L$, respectively. $ i, j = 1, 2, 3$ denote the generations. 
All the chiral superfields include the left-handed fermions in the flavor basis. $\vph_i$ and $V_{ij}$ correspond to  additional phases in the minimal SUSY GUT and the CKM matrix with the constraint $\sum_i \vph_i=0$.
$Q$ and $L$ denote the weak-doublet chiral superfields for left-handed quarks and left-handed leptons, respectively:
\eqs{
Q_i = \left( 
\begin{array}{c}
U_i \\
V_{ij} D_j
\end{array} \right), ~~~~~~~
L_i = \left( 
\begin{array}{c}
N_i \\
E_i
\end{array} \right),
}
where $U, D, E$, and $N$ are the chiral superfields for left-handed up-type and down-type quarks, and left-handed charged and neutral leptons, respectively. 
$U^C, D^C$, and $E^C$ denote the chiral superfields for the charge-conjugation of right-handed up-type and down-type quarks, and right-handed charged lepton, respectively.
In the Higgs sector, there are $\bold 5$, $\ovl{\bold 5}$, and $\bold{24}$ representational superfields,
\eqs{
H^A_{\bold 5}(\bold 5) & = \left( 
\begin{array}{c}
H_C^\alpha \\
H_u^r 
\end{array} \right), ~~~ 
H_{\ovl{\bold 5} A}  (\bold{\bar 5}) = \left( 
\begin{array}{c}
H_{\ovl C\alpha} \\
\ep_{rs} H_d^s 
\end{array} \right), \\
{\Si^A}_B(\bold{24}) & =\left ( 
\begin{array}{cc}
\Si_8 & \Si_{(3,2)} \\
\Si_{(3^\ast,2)} & \Si_3
\end{array}
\right ) +  \frac{1}{\sqrt{60}} \left ( 
\begin{array}{cc}
2 & 0 \\
0 & -3
\end{array}
\right ) \Si_{24}.
}
$H_{\bold 5}(\bold 5)$ and $H_{\ovl{\bold 5}} (\ovl{\bold 5})$ include the MSSM Higgs doublets, $H_u$ and $H_d$. In order to embed the MSSM Higgs multiplets in the $SU(5)$ multiplets, we have to introduce the color-triplet Higgs multiplets $H_C$ and $H_{\ovl C}$. The adjoint Higgs multiplet $\Sigma(\bold{24})$ is introduced to cause the spontaneous symmetry breaking of the $SU(5)$ gauge symmetry according to the non-zero vacuum expectation value (VEV) of $\Si_{24}$.

The Lagrangian for the minimal SUSY $SU(5)$ GUT is given by
\eqs{
{\cal L} = \int d^4\theta ~ {\cal K}_{\rm MSGUT} + \left[ \int d^2\theta ~ \left( W_{\rm MSGUT} + \frac{1}{8g_5^2} \tr {\cal W}^{\alpha} {\cal W}_\alpha \right) + {\rm h.c.} \right] + {\cal L}_{\text{ghost}} + {\cal L}_{\text{gauge-fixing}},
} 
where ${\cal K}_{\rm MSGUT}$ and $W_{\rm MSGUT}$ are the K\"ahler potential and the superpotential, respectively. $g_5$ denotes the unified gauge coupling constant.
The field strength chiral superfield ${\cal W}^\alpha$ consists of vector superfield ${\cal V}_5={\cal V}_5^A T^A$, where $T^A$ is the generator of $SU(5)$ with $\tr T^A T^B = \frac12 \del^{AB}$:
\eqs{
{\cal W}^\alpha = - \frac14 \ovl \coD^2 (e^{-2g_5 {\cal V}_5} \coD^\alpha e^{2g_5 {\cal V}_5}).
\label{eq:field_strength}
}
Here, $\coD$ and $\ovl \coD$ denote the covariant derivatives on superspace. 
The vector superfield ${\cal V}_5$ is decomposed in terms of the SM gauge group:
\eqs{
{\cal V}_5 =  \left(
\begin{array}{cc}
\dps {G^\alpha}_\beta - \frac{2}{\sqrt{60}} B{\del^\alpha}_\beta & \dps \frac{1}{\sqrt 2} {X^{\dag \alpha}}_r \vspace{+.5em}\\
\dps \frac{1}{\sqrt 2} {X^s}_\beta & \dps {W^s}_r +  \frac{3}{\sqrt{60}} B {\del^s}_r
\end{array}
\right). \label{eq:decomp_V}
}
$G, W$, and $B$ are the vector superfields for $SU(3)_C, SU(2)_L$, and $U(1)_Y$, respectively, and they are defined as
\eqs{
{G^\alpha}_\beta = G^a {(T^a)^\alpha}_\beta, ~~~~~~
{W^s}_r  = W^a {(t^a)^s}_r,
}
using the generators $T^a$ and $t^a$ of $SU(3)_C$ and $SU(2)_L$, respectively.
$X$ is the vector superfield for the $X$ boson, which induces baryon-number violating operators. It acquires the heavy mass by eating the Nambu-Goldstone (NG) modes, $\Si_{(3,2)}$ and $\Si_{(3^\ast,2)}$, after the $SU(5)$ symmetry breaking. $M_X$ denotes the mass for the $X$ boson in this paper.

In the minimal SUSY $SU(5)$ GUT, the K\"ahler potential and the superpotential in the flavor basis of matter superfields are written as
\eqs{
{\cal K}_{\rm MSGUT} & = \Phi^{\dag A}_i {(e^{-2g_5{\cal V}_5})^B}_A \Phi_{iB} 
+ \Psi^\dag_{i AB} {(e^{2g_5{\cal V}_5})^A}_C {(e^{2g_5{\cal V}_5})^B}_D \Psi^{CD}_i  \\ 
& ~~~~~ + 2 {\Si^{\dag A}}_B {(e^{-2g_5{\cal V}_5})^C}_A {(e^{2g_5{\cal V}_5})^B}_D {\Si^D}_C 
+ H^{\dag A}_{\ovl{\bold 5}} {(e^{-2g_5{\cal V}_5})^B}_A H_{\ovl{\bold 5} B} \\
& ~~~~~ + H^\dag_{\bold{5} A} {(e^{2g_5{\cal V}_5})^A}_B H^B_{\bold 5}, \\
\\
W_{\rm MSGUT}& = \ \frac{y}{3} \tr \Si^3 + \frac{y v_\Si}{2} \tr \Si^2 + y_H H_{\ovl{\bold{5}}A} (\Si^A_B + 3 \mu_0 \del^A_B) H_{\bold{5}}^B  \\
& ~~~~~ + \frac{y_u^i}{4} e^{i\vph_i} \ep_{ABCDE} \Psi^{AB}_i \Psi^{CD}_i H_{\bold{5}}^E + \sqrt 2 V^\ast_{ij} y_d^j \Psi^{AB}_i \Phi_{j A} H_{\ovl{\bold{5}}B}.
}
$y$ denotes the cubic coupling constant of the adjoint Higgs multiplet and $v_{\Si}$ is the VEV of the adjoint Higgs multiplet. $y_u^i$ and $y_d^i$ denote the diagonalized Yukawa matrices.

The adjoint Higgs multiplet and the color-triplet Higgs multiplets acquire heavy masses through the interactions in the superpotential. 
The doublet-triplet splitting is achieved by tuning $\mu_0$ in the minimal SUSY $SU(5)$ GUT.
$M_{H_C}(=5 y_H v_{\Si})$ denotes the mass of the color-triplet Higgs multiplets. The masses of the adjoint Higgs multiplets are also split after the $SU(5)$ symmetry breaking. The triplet $\Si_3$ and the octet $\Si_8$ have a common mass denoted as $M_\Si(= \frac52 y  v_{\Si})$, and the mass for $\Si_{24}$ is $M_{\Si_{24}}(= \frac12 y  v_{\Si})$. The $X$ boson mass $M_X$ is $5\sqrt2 g_5  v_{\Si}$. Note that $y$ and $y_H$ should be large, if the color-triplet Higgs and adjoint Higgs multiplets are much heavier than the $X$ boson.

In the minimal setup of the SUSY $SU(5)$ GUT, the $X$-boson interactions with the matter superfields are given by
the following terms,
\begin{eqnarray}\label{eq:x_boson_interaction}
{\cal L}_X &=& \int d^4\theta \left( {\cal K}^{(0)}_{V_1} + {\cal K}^{(0)}_{V_2}+ {\cal K}^{(0)}_{V_3} \right ),  \\
&&{\cal K}^{(0)}_{V_1} =- \sqrt 2 g_5   \ep^{rs} L^\dag_{si} D^C_{\alpha i} {X^\dag}_r^\alpha + {\rm h.c.}, \\
&&{\cal K}^{(0)}_{V_2} =- \sqrt 2 g_5 \ep_{\alpha\beta\gamma} e^{i\vph_i}U^{C \dag \gamma}_i Q^{r \beta}_i {X^\dag}^\alpha_r + {\rm h.c.}, \\
 &&{\cal K}^{(0)}_{V_3} =- \sqrt 2 g_5\ep^{sr}V_{ij} Q^\dag_{s\alpha i} E^C_j {X^\dag}^\alpha_r + {\rm h.c.}, 
 \end{eqnarray}
and the baryon-number violating operators are effectively induced by integrating out the $X$ boson at the low energy. The effective dimension-six operators are written as follows at the tree level;
\footnote{Notice that 
the propagators of the vector superfields differ from those of canonically normalized gauge bosons by a factor $1/2$ under our convention for the kinetic terms of the vector superfields.} 
\begin{eqnarray}
{\cal L}_{\text{dim.6}} &=  & \int d^4 \theta \left ( {\cal K}^{(0)}_1+{\cal K}^{(0)}_2 \right )   \label{eq:dim_6} \\
&&{\cal K}^{(0)}_1= - e^{i\vph_i} \frac{g_5^2}{M_X^2} \ep_{\alpha\beta\gamma} \ep_{rs}   U^{C \dag \alpha}_i D_j^{C\dag \beta}  Q_i^{r \, \gamma} L_j^s + {\rm h.c.} \\
&&{\cal K}^{(0)}_2=  -e^{i\vph_i} V_{kj}^\ast \frac{g_5^2}{M_X^2} \ep_{\alpha\beta\gamma} \ep_{rs}  E_j^{C\dag} U^{C \, \alpha \, \dag}_i  Q_k^{\beta r} Q_i^{s \gamma} + {\rm h.c.}.
\end{eqnarray}
Below, we investigate the one-loop correction to the $4$-Fermi interactions
and especially estimate how large the threshold correction is according to the heavy particles decoupling around the GUT scale.
We focus on the operators relevant to nucleon decay in not only the minimal SUSY $SU(5)$ GUT but also its vector-like extensions,
where $SU(5)$ vector-like chiral superfields are additionally introduced. 
In the later case, we simply assume that the vector-like pairs have supersymmetric masses without the mixing between the extra fields and the MSSM fields. 
We only discuss the gauge interactions in our calculation. The gauge interactions in the minimal SUSY SU(5) GUT, which are relevant to the evaluation of the threshold correction to the baryon-number violating operators, are summarized in \cref{ap:gauge int}. For simplicity, we omit the generation indices ($i, j \cdots$) below.

\section{Radiative Correction to the  Baryon-Number Violating Operators\label{sec:radiative}}
In the supersymmetric theories, effective K\"ahler potentials are useful to derive the radiative corrections. 
In order to evaluate the corrections to the baryon-number violating dimension-six operators induced by the $X$ boson,
we discuss the effective K\"ahler potentials at the one-loop level, and evaluate the threshold corrections to the operators.

First of all, let us discuss a general effective supersymmetric action $\Gamma[\Phi, \Phi^\dag]$,
which is the function of chiral superfield $\Phi$, antichiral superfield $\Phi^\dag$, and their derivatives. The general form of the effective supersymmetric action would be as follows,
\eqs{
\Gamma[\Phi, \Phi^\dag] = & \int d^4 xd^4 \theta {\cal L}_{\rm eff}(\Phi, \coD_A \Phi, \coD_A \coD_B \Phi, \cdots, \Phi^\dag, \coD_A \Phi^\dag, \coD_A \coD_B \Phi^\dag, \cdots) \\
& + \left\{ \int d^4x d^2\theta {\cal L}^{(c)}_{\rm eff}(\Phi, \coD_A \Phi, \coD_A \coD_B \Phi, \cdots) + {\rm h.c.} \right \},
}
where $\coD_A$ is the superspace covariant derivative which consists of $\partial_\mu$, $\coD_\alpha$, and $\ovl \coD_{\dot \alpha}$. Here, we do not include vector superfields for simplify. The perturbative corrections appear only in the $D$ term  due to the non-renormalization theorem. 
The effective supersymmetric Lagrangian ${\cal L}_{\rm eff}$ is divided into two parts under ${\partial_\mu \Phi = 0}$,
\eqs{
{\cal L}_{\rm eff} = {\cal K}(\Phi, \Phi^\dag) + {\cal F}(\coD_\alpha \Phi, \coD^2 \Phi, \ovl \coD_{\dot\alpha} \Phi^\dag, \ovl \coD^2 \Phi^\dag ;\Phi, \Phi^\dag),
}
where ${\cal K}(\Phi, \Phi^\dag)$ is the effective K\"ahler potential and ${\cal F}(\coD_\alpha \Phi, \coD^2 \Phi, \ovl \coD_{\dot\alpha} \Phi^\dag, \ovl \coD^2 \Phi^\dag ;\Phi, \Phi^\dag)$ is called the effective auxiliary potential. While some diagrams may generate the terms including superfields on which more than three covariant derivatives act, we may always obtain the above form by using algebra of super-covariant derivatives ($\coD$ algebra). 
The effective auxiliary potential vanishes in the limit that $\coD_\alpha \Phi =0$ and $\ovl \coD_{\dot\alpha} \Phi^\dag = 0$,
so that the effective K\"ahler potential is identified by taking the limit.

Below, we study the threshold corrections to the baryon-number violating dimension-six operators at the GUT scale with the effective K\"ahler potential.  First, we calculate the effective actions for constant fields in both full and effective theories at the one-loop level with the supergraph technique \cite{Grisaru:1979wc}.  
We adopt the modified dimensional reduction ($\ovl{\text{DR}}$) scheme \cite{Siegel:1979wq} as the renormalization scheme of the gauge coupling constants while we impose the on-shell condition for the $X$ boson mass.  
We also introduce the IR cut off in order to control fictitious IR singularities.  Then, we identify the effective K\"ahler potential for the baryon-number violating operators by taking $\coD_\alpha \Phi =0$ and $\ovl \coD_{\dot\alpha} \Phi^\dag = 0$ together with the $\coD$ algebra. 
By matching the effective K\"ahler potentials in full and effective theories, 
we derive the one-loop threshold corrections to the Wilson corrections of the dimension-six operators.

\subsection{Radiative Corrections in the Full Theory}
In this subsection, we show the radiative corrections to the baryon-number violating dimension-six operators in the full theory, where the $X$ boson is activated. The radiative corrections consist of  the wave function renormalization of quarks and leptons, the vacuum polarization of the $X$ boson, the vertex correction, and the box-like corrections. In this section, we show only the results of the supergraph calculation. Details of the calculations are given in \cref{app:explicit_loops}.

\subsubsection*{Two-Point Functions for Matter Fields}

First we study two-point functions for matter fields at the one-loop level.
The functions generally include UV divergences which are renormalized by the wave function renormalization factors. 
We estimate the factors in the $\ovl{\text{DR}}$ scheme, ignoring the contributions from the Yukawa interactions. 
The radiative corrections to the two-point functions via the gauge interactions are determined by the gauge groups,
in the both of the full theory and the EFT. 

In general, the one-loop renormalized two-point function for chiral superfield $\Phi$ is defined as $\Gamma^{\text{2-pt}}_\Phi = \widetilde\Gamma^{\text{2-pt}}_\Phi + Z_\Phi -1$.
The wave function renormalization constant for the matter superfield $Z_\Phi$ absorbs  the UV divergent terms proportional to $1/{\ep'}$ in the $\ovl{\text{DR}}$ scheme: \footnote{${2}/{\ep'} \equiv {2}/{\ep} - \gamma + \ln 4\pi$ is defined  and $\ep$ satisfies $\ep = 4 - d$ in the $d$-dimension momentum space.}
\begin{equation}
Z_\Phi = 1 + c^{\Phi}_5 \frac{g_5^2}{4\pi^2} \times \frac{1}{\ep'},~
Z^{\rm EFT}_\Phi = 1 + \sum^3_{n=1} c^{\Phi}_n   \frac{g_n^2}{4\pi^2} \times \frac{1}{\ep'}.
\end{equation}
$Z_\Phi$ and $Z_\Phi^{\rm EFT}$ denote the wave function renormalization factors in the full  and the effective theories, respectively. $g_3, g_2$, and $g_1$ are the gauge couplings of $SU(3)_C, SU(2)_L$, and unified $U(1)_Y$ gauge symmetries. $c^{\Phi}_5$ and $c^{\Phi}_n$ ($n=3,\,2,\,1$) are the quadratic Casimir of $\Phi$
in $SU(5)$, $SU(3)_C$, $SU(2)_L$, and GUT normalized $U(1)_Y$ gauge symmetries.\footnote{ $c^{\Phi}_1$ is given by $c^{\Phi}_1=(Q^{\Phi}_Y)^2 \times (3/5)$, where $Q^{\Phi}_Y$ is hypercharge of $\Phi$.}

Then, the one-loop renormalized two-point function in the full theory is given by
\eqs{
\Gamma^{\text{2-pt;1-loop}}_{\Phi} = \left( 1 + a_{\Phi} f(M_X^2) + b_{\Phi} f(\mu_{\rm IR}^2) \right) \Gamma^{\text{2-pt}}_{\Phi \, 0}.
 \label{eq:fff}}
$\Gamma^{\text{2-pt}}_{\Phi \, 0}$ is the tree-level two-point function, and
$a_\Phi$ and $b_\Phi$ are the constants obtained from the one-loop calculations,
\eqs{
a_{\Phi}= - (c^{\Phi}_5 - \sum^3_{n=1} c^{\Phi}_n) \frac{g_5^2}{8\pi^2},&~~~~
b_{\Phi}=-  \sum^3_{n=1} c^{\Phi}_n \frac{g_5^2}{8\pi^2}.
}
We set the mass of the MSSM vector superfields to be a non-zero value which is denoted by $\mu_{\rm IR}$ in order to regularize the IR divergence, as mentioned above. The function $f$ in \cref{eq:fff} is defined as 
\eqs{
f(M^2) & \equiv 
1 - \ln\frac{M^2}{\mu^2},  \label{eq:fcs_wr}
}
where $\mu$ denotes the renormalization scale in the $\ovl{\text{DR}}$ scheme. The two-point function in the effective theory is derived by removing the $X$ boson contribution in \cref{eq:fff} when $g_5=g_3=g_2=g_1$.

\begin{figure}[t]
\begin{center}

\includegraphics[width=7cm,clip]{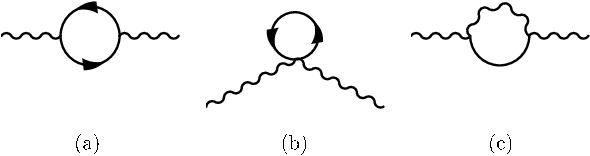}
\caption{Diagrams of chiral multiplets for
radiative correction to two-point function of superfield for $X$ boson.}
\label{fig:quad}
\end{center}
\end{figure}

\subsubsection*{Vacuum Polarization}

Next, we estimate the radiative corrections to the propagator for the $X$ boson. 
Not only the MSSM fields but also the GUT-scale fields such as the $SU(5)$-adjoint field  
contribute to the vacuum polarization of the $X$ boson.

The chiral superfields have three kinds of the contributions which are described in \cref{fig:quad}. The diagrams (a) and (b) are induced by the supergauge interaction $\Phi^\dag V \Phi$ and $\Phi^\dag V^2 \Phi$, respectively. The diagram (c) is generated by the $SU(5)$-breaking adjoint Higgs superfield, which has interactions $\langle\Si^\dag\rangle V^2 \Si$ and 
$\Si^\dag V^2 \langle\Si\rangle$ after acquiring the VEV. 
\begin{figure}[t]
\begin{center}
\includegraphics[width=7cm,clip]{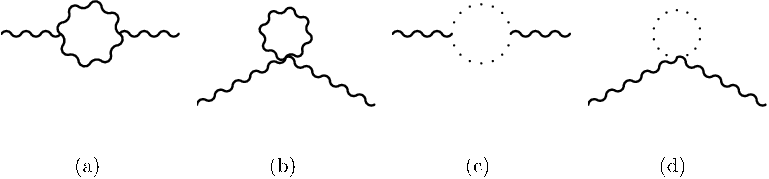}
\caption{Diagrams
of gauge and ghost superfields for radiative correction to two-point function of superfield for $X$ boson.}
\label{fig:quad2}
\end{center}
\end{figure}

For the gauge sector, we have the four-type diagrams to contribute to the two-point function of the $X$ boson. The diagrams (a) and (b) in \cref{fig:quad2} arise from the self interactions of the vector superfields. If the internal vector superfields in the diagram (b) are massless, the diagrams have  no contribution to the two-point function in the $\ovl{\text{DR}}$ scheme. The diagrams (c) and (d)  show the ghost loop contribution. 

Finally, the two-point function of the $X$ boson is in the form as below:
\eqs{
\Gamma_X^{(2)}(k^2) = k^2 - M_X^2 - \Si_X(k^2),
}
where $\Si_X(k^2)$ is the renormalized vacuum polarization for $X$ boson. The UV divergence
in the one-loop corrections 
is absorbed by the wave function factor ($Z_X$) and mass $(M_X)$ of the $X$ boson. In this paper the on-shell condition for the $X$ boson mass is imposed so that this leads the equation $\Si_X(M_X^2)=0$. This is because heavy particles are decoupled from $\Si_X(0)$ under the on-shell condition, if they have $SU(5)$ symmetric masses much larger than  the $X$ boson mass.\footnote{
The GUT-scale mass spectrum may be constrained using the gauge coupling unification \cite{Hisano:1992mh,Hisano:1992jj}. In the works, they use the threshold correction to the gauge coupling constants at the GUT scale at the one-loop level so that the renormalization condition for the $X$ boson mass does not appear there. We need the threshold correction at the two-loop level in order to get the constraint on the on-shell $X$ boson mass.
}
$\Si_X(0)$ will appear in the threshold correction to the baryon-number violating operators.

The counter term $\del Z_X$ is determined to absorb the UV divergence which arise from the gauge contributions and the matter contributions such as \cref{fig:quad,fig:quad2}. We obtain 
\eqs{
\del Z_X = Z_X - 1 = \frac{g_5^2}{8\pi^2} \left ( 3 C_2(G) - \sum_R T(R)\right ) \times \frac{1}{\ep'},
}
where $T(R) \del^{ab} = \tr (T_R^a T_R^b)$ and $C_2(G) \del_i^j= \sum_a (T_G^aT_G^a)_i^j$ are defined.  As expected, $\del Z_X$ is proportional to the one-loop beta function for the $SU(5)$ gauge coupling constant. In the SUSY $SU(5)$ GUT models with $\bold 5 + \ovl{\bold 5}$ vector-like matter superfields and $\bold{10} + \ovl{\bold{10}}$ vector-like matter superfields, we find
\eqs{
\sum_R T(R) = \frac12 (N_f+2+2n_{\bold 5}) + \frac32 (N_f + 2n_{\bold{10}}) + 5,
}
where $N_f, n_{\bold 5}$, and $n_{\bold{10}}$ are the number of generations, $\bold 5 + \ovl{\bold 5}$ and $\bold{10} + \ovl{\bold{10}}$ vector-like pairs, respectively. 

In the SUSY $SU(5)$ GUT with extra vector-like matters, the vacuum polarization $\Si_X(p^2)(={\ovl\Si_X(p^2)}-{\ovl\Si_X(M_X^2)})$ is given by 
\eqs{
 \frac{16\pi^2}{2g_5^2} {\ovl \Si}_X(p^2) 
= & \left[ \frac12 (N_{\bold{5}} + N_{\ovl{\bold{5}}}) + \frac32 (N_{\bold{10}} + N_{\ovl{\bold{10}}}) \right] B(p^2, 0, 0)
+ \frac{25}{6} B(p^2, M_\Si^2, M_X^2) \\
& + \frac{5}{6} B(p^2, M_{\Si_{24}}^2, M_X^2) +  B(p^2, M_{M_{H_C}}^2, 0) \\
& + \frac{5}{12} M_X^2 \left[ 3A(p^2,M_X^2,0) + 10 A(p^2,M_X^2,M_\Si^2) + 2 A(p^2,M_X^2,M_{\Si_{24}}) \right]\\
& - 5 C_2(G) \frac{p^2}{4} A(p^2,M_X^2,0) - \frac12 C_2(G) B(p^2,M_X^2,0) \\
& + (p^2\text{-independent terms})
,
\label{eq:vacpol}}
where $N_{r}$ $(r=\bold{5},{\ovl{\bold{5}}},{\bold{10}},{\ovl{\bold{10}}})$
 denotes the number of the massless superfields in $r$ representation. The loop functions $A$ and $B$ are defined as 
\eqs{
A(p^2,M_1^2,M_2^2) & \equiv \int^1_0 dx  \ln \frac{\Delta}{\mu^2}, \\
B(p^2,M_1^2,M_2^2) & \equiv \int^1_0 dx \left[ \Delta- (2\Delta + x(x-1)p^2 ) \ln \frac{\Delta}{\mu^2} \right], \label{eq:loop_func}
}
where $\Delta = x(x-1) p^2 + x M_2^2 + (1-x) M_1^2$ is defined. 

The first and second lines in \cref{eq:vacpol} correspond to the contributions of the massless and massive fields in \cref{fig:quad}(a).  
The third line is for diagram (c) in \cref{fig:quad}, in which the VEV of the adjoint Higgs multiplet is included in the vertices. In the fourth line, we show the contributions from the gauge sector: The first term in the forth line is induced by the three-vector interactions (\cref{fig:quad2}(a)), while the second term corresponds to the ghost diagrams (\cref{fig:quad2}(c)).  
The $p^2$- independent terms come from the diagrams \cref{fig:quad}(b), \cref{fig:quad2}(b), and \cref{fig:quad2}(d). 

We finally obtain the full one-loop corrections to the two-point function by summing of the contributions from the chiral superfields, the vector superfield, and the ghost superfields. The resumed propagator $D_{XX}(p^2)$ of $X$ superfield in terms of the superfield notation is given by $D_{XX}(p^2) =-i/(2\Gamma_X^{(2)}(p^2))$.

After the spontaneous symmetry breaking of the GUT gauge symmetry, the baryon-number violating dimension-six operators are  induced by  the $X$ boson, and the coefficients are proportional to $1/M_X^2$. In order to match the full and the effective theories at the one-loop level, we need to take into account the one-loop corrections to the propagator of the $X$ boson. Since the momenta of external fields in the baryon-number violating dimension-six operators are negligible compared with the $X$ boson mass, we may set the momentum of internal $X$ boson zero.

\subsubsection*{Vertex Corrections}

Next, we show the one-loop vertex corrections to the $X$ boson interactions with quarks and leptons. The tree-level interactions are given in \cref{eq:x_boson_interaction}.

\begin{figure}[t]
\begin{center}

\includegraphics[width=8cm,clip]{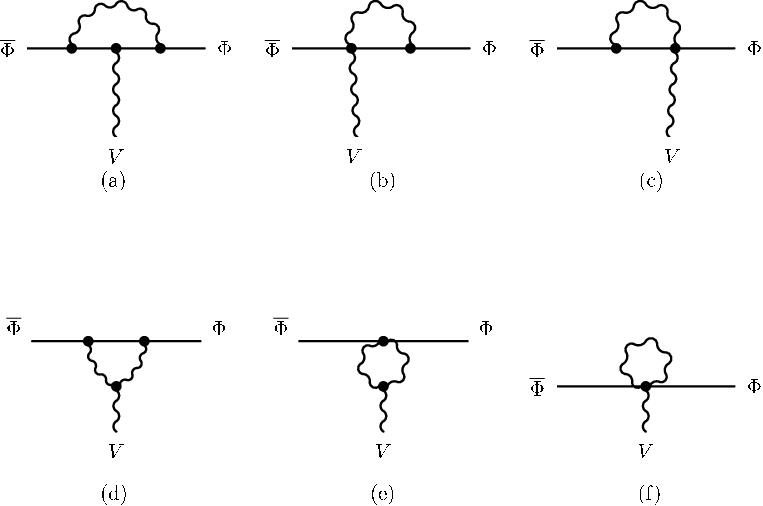}
\caption{Diagrams for vertex correction}
\label{fig:ver_corr}
\end{center}
\end{figure}

Several one-loop diagrams in \cref{fig:ver_corr} contribute to the vertex corrections. Since the supersymmetric gauge interactions in terms of the superfield formalism have the form $\Phi^\dag e^{2g V} \Phi$ ($\Phi$ is a matter chiral superfield, and $V$ and $g$ are a vector superfield and its gauge coupling, respectively), there exist diagrams which do not appear in component calculation. The diagram (a) has only the vertex $2g \Phi^\dag V \Phi$, and the diagrams (b) and (c) include the vertex $2g^2 \Phi^\dag V^2 \Phi$. The diagrams (d) and (e) include the three-point self interactions of vector superfields. Since the external vector superfield is for the broken gauge symmetry, two internal vector superfields must be massive and massless ones.
The contribution from the diagram (f) is vanishing due to the superspace integral.

Thus, we calculate the contributions from the diagrams (a)-(e) in \cref{fig:ver_corr}. The momenta of all the external superfields are set to be $p^2=0$, for simplicity. In some diagrams, since they contain IR divergent contributions in this momentum assignment, the non-zero masses of the MSSM vector superfields ($\mu_{\rm IR}$) are introduced as IR regulators.  
Under this momentum assignment, we carry out loop momentum integrals and Grassmann integrals, and we discard the auxiliary terms.  
We expand the one-loop K\"ahler terms around $p^2=0$, and then we extract the dominant contributions around $p^2=0$.
The vertex corrections to the gauge interactions between the MSSM matter fields and the $X$ boson are as follows:
\eqs{
{\cal K}^{(1)}_{V_1} & = \left[
- \frac25 C_{1}^{\rm (v)}(\mu_{\rm IR}) 
+ \frac{21}{5} C_{2}^{\rm (v)}(\mu_{\rm IR}) 
+ 5 C_{2}^{\rm (v)}(M_X)  \right]{\cal K}^{(0)}_{V_1}, \\
{\cal K}^{(1)}_{V_2} & = \left[ 
\frac{12}{5} C_{1}^{\rm (v)}(\mu_{\rm IR}) 
- 2 C_{1}^{\rm (v)}(M_X) 
+ \frac{49}{5} C_{2}^{\rm (v)}(\mu_{\rm IR}) 
+ 9 C_{2}^{\rm (v)}(M_X) \right] {\cal K}^{(0)}_{V_2}, \\
{\cal K}^{(1)}_{V_3} & =
\left[
 \frac{2}{5} C_{1}^{\rm (v)}(\mu_{\rm IR}) 
- 4 C_{1}^{\rm (v)}(M_X) 
+ \frac{29}5 C_{2}^{\rm (v)}(\mu_{\rm IR}) 
+ 13 C_{2}^{\rm (v)}(M_X)  \right] {\cal K}^{(0)}_{V_3}.
}
The contributions from the diagrams (d) and (e) in \cref{fig:ver_corr}  are canceled each other.
The coefficients $C_{1}^{\rm (v)}$ and $C_{2}^{\rm (v)}$ correspond to the correction from the diagram (a), and the ones from the diagrams (b) and (c) in \cref{fig:ver_corr}, respectively.
After the loop momentum and superspace integrals, we find that $C_{1}^{\rm (v)}$ and $C_{2}^{\rm (v)}$  are given by the functions
of the mass of the internal vector superfield $M$, 
\eqs{
C_{1}^{\rm (v)}(M) = - C_{2}^{\rm (v)}(M) \equiv
 \frac12 \frac{g_5^2}{16 \pi^2}\left[ \frac2{\ep'} + 1 - \ln \frac{M^2}{\mu^2} \right].
}
These loop functions are the coefficients of the effective K\"ahler potential ${\cal K}_{1}^{\rm (v)}$ and ${\cal K}_{2}^{\rm (v)}$ defined in \cref{app:explicit_loops} in the limit that $p^2$ vanishes.

Now, we determine the renormalization constants for the vertices. One-loop renormalized vertex functions are given by
\eqs{
{\cal K}_{V_1} & = {\cal K}^{(0)}_{V_1} + {\cal K}^{(1)}_{V_1} + \left( Z_L^{1/2} Z_D^{1/2} Z_X^{1/2} Z_{{\cal C}_{V_1}} - 1\right) {\cal K}^{(0)}_{V_1}, \\
{\cal K}_{V_2} & = {\cal K}^{(0)}_{V_2} + {\cal K}^{(1)}_{V_2} + \left( Z_U^{1/2} Z_Q^{1/2} Z_X^{1/2} Z_{{\cal C}_{V_2}} - 1\right) {\cal K}^{(0)}_{V_2} ,  \\
{\cal K}_{V_3} & = {\cal K}^{(0)}_{V_3} + {\cal K}^{(1)}_{V_3} + \left( Z_Q^{1/2} Z_E^{1/2} Z_X^{1/2} Z_{{\cal C}_{V_3}} - 1 \right) {\cal K}^{(0)}_{V_3}. \label{eq:ver_corr}
}
When ${\cal K}_{V_n}$~($n=1$, $2$, $3$) are described as ${\cal K}_{V_n}={\cal C}_{V_n} {\cal O}_{V_n}$ with the operators $ {\cal O}_{V_n}$ and the Wilson coefficients ${\cal C}_{V_n}$,
$Z_{{\cal C}_{V_n}}$ are defined to renormalize the UV divergences in ${\cal C}_{V_n}$.
Then we find
\eqs{
Z_{{\cal C}_{V_1}} = Z_{{\cal C}_{V_2}}  = Z_{{\cal C}_{V_3}} = 1 - \frac{g_5^2}{16\pi^2} \left( 3 C_2(G) - \sum_R T(R)\right) \times \frac{1}{\ep'},
}
which are consistent with the one-loop beta function for the gauge coupling.

\subsubsection*{Box-like Corrections}

\begin{figure}[t]
\begin{center}

\includegraphics[width=7cm,clip]{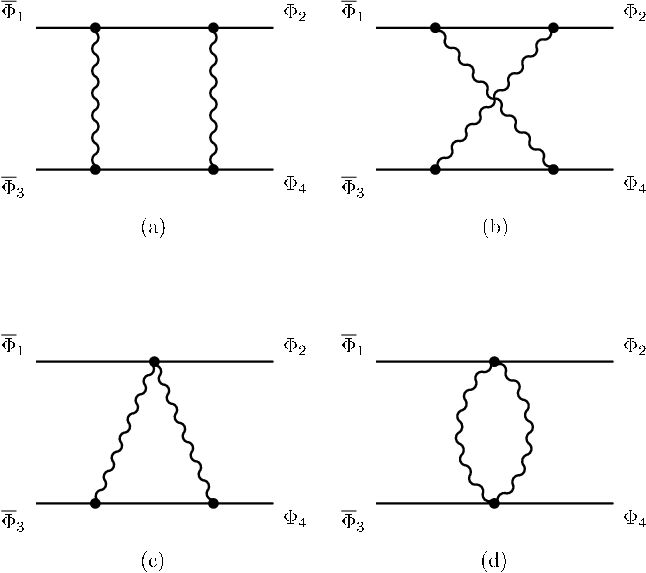}
\caption{Box-like diagrams}
\label{fig:box_corr}
\end{center}
\end{figure}

The box-like diagrams contribute to the radiative corrections of the dimension-six operators. \cref{fig:box_corr} shows all type of the box-like diagrams; we refer to the diagram (a) as the box diagram, the diagram (b)  as the crossing box diagram, and the diagram (c)  as the triangle diagram. The diagram (d) vanishes due to the superspace integral.
Thus, it is sufficient that we evaluate the diagrams (a)-(c) in \cref{fig:box_corr}. In these figures, one of two internal gauge superfield lines must be massive since we focus on the baryon-number violating operators. As is the case in the vertex corrections, we set all momenta of the external superfields to be $p^2=0$ and the fictitious masses of the MSSM vector superfields to be $\mu_{\rm IR}$, and we remove the auxiliary terms. 
For the momentum assignment, we find that the box diagram (a) vanishes while the crossing box diagram (b) and the triangle diagram (c) are given by the following functions:
\eqs{
C_{\rm cross}(M_X) = - C_{\rm triangle}(M_X) \equiv - \frac{1}{2} \frac{g_5^2}{16\pi^2} \ln \frac{M_X^2}{\mu_{\rm IR}^2}.
}
These loop functions correspond to the coefficients in the effective K\"ahler potential ${\cal K}_{\rm cross}$ and ${\cal K}_{\rm triangle}$ defined in \cref{app:explicit_loops} in the limit that $p^2$ vanishes.

In SUSY $SU(5)$ GUTs, the baryon-number violating dimension-six operators are generated at the tree level in 
\cref{eq:dim_6}.
The one-loop radiative corrections from the box-like diagrams are written by $ C_{\rm cross}$ and $C_{\rm triangle}$:
\eqs{
{\cal K}_{1}^{\text{(Box)}} & = \left[
 - \frac{18}{5} C_{\rm cross}(M_X) 
 + \frac{14}{5} C_{\rm triangle}(M_X)\right] {\cal K}_{1}^{(0)}, \\
{\cal K}_{2}^{\text{(Box)}} & = \left[
 - \frac{14}{5} C_{\rm cross}(M_X)
 + \frac{22}{5} C_{\rm triangle}(M_X)\right] {\cal K}_{2}^{(0)}.
}

\subsection{Radiative Corrections in EFT}
\begin{figure}[t]
\begin{center}
\includegraphics[width=10cm,clip]{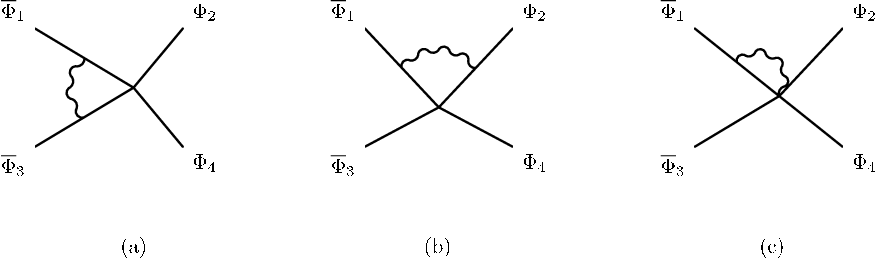}
\caption{Radiative corrections in EFT}
\label{fig:sd-eff}
\end{center}
\end{figure}

Now we consider the radiative correction to the higher-dimensional K\"ahler terms in the EFT. There are three kinds of contributions to the radiative correction. The first one is the diagram (a) in \cref{fig:sd-eff}, where a vector superfield is attached to two chiral superfields or two antichiral superfields. The second is the diagram (b),  in which a vector superfield is attached to both a chiral and an antichiral superfield. The third one is the radiative corrections induced by the gauge interaction of the composite operators. 

We adopt the same momentum assignment which we used in the full theory. After the loop momentum and the superspace integrals,  we derive the one-loop corrections as 
\eqs{
{\cal K}_{1}^{(1):\text{eff}} & =\left[ 
\left( \frac{16}{3} g_3^2 +\frac{4}{15} g_1^2  \right) C^{\rm EFT}_1 (\mu_{\rm IR}) 
+ \left( \frac{32}{3}g_3^2+\frac{8}{15} g_1^2 \right) C^{\rm EFT}_2 (\mu_{\rm IR}) \right] {\cal K}_{1}^{(0)}, \\
{\cal K}_{2}^{(1):\text{eff}} & =\left[ 
\left( \frac{16}{3} g_3^2 +\frac{4}{15} g_1^2  \right) C^{\rm EFT}_1 (\mu_{\rm IR}) 
+ \left( \frac{32}{3}g_3^2+\frac{8}{15} g_1^2 \right) C^{\rm EFT}_2 (\mu_{\rm IR}) \right] {\cal K}_{2}^{(0)}. \\
}
Here, the diagram (a) vanishes while the diagrams (b) and (c) are given by 
$C^{\rm EFT}_1 (\mu_{\rm IR})$ and $C^{\rm EFT}_2 (\mu_{\rm IR})$, respectively:
\eqs{
C^{\rm EFT}_1 (\mu_{\rm IR}) & = -C^{\rm EFT}_2 (\mu_{\rm IR}) 
\equiv \frac{1}{16\pi^2} \frac{1}{2} \left( \frac{2}{\ep'} + 1 - \ln \frac{\mu_{\rm IR}^2}{\mu^2} \right). \\
}
These functions correspond to the coefficients defined in \cref{eq:kahler_EFT} in the limit: $p^2$ vanishes.
The effective K\"ahler potentials up to the one-loop level are described as   
\eqs{
{\cal K}^\text{eff}_{1} & = {\cal K}^\text{(1):eff}_{1} + {\cal K}^{(0)}_{1} 
+ \left( Z_{{\cal C}_{1}} {Z^{\text{EFT}}_U }^{1/2}{Z^{\text{EFT}}_Q }^{1/2}{Z^{\text{EFT}}_D }^{1/2}{Z^{\text{EFT}}_L }^{1/2} -1 \right) {\cal K}^{(0)}_{1}, \\
{\cal K}^\text{eff}_{2} & = {\cal K}^\text{(1):eff}_{2} + {\cal K}^{(0)}_{2} 
+ \left( Z_{{\cal C}_{2}} {Z^{\text{EFT}}_E }^{1/2}{Z^{\text{EFT}}_U }^{1/2}Z^{\text{EFT}}_Q  -1 \right) {\cal K}^{(0)}_{2}.
}
The logarithmic divergences are absorbed by the counter terms of ${\cal C}_{A}$, and then we have
\eqs{
Z_{{\cal C}_{1}} & = 1 - \frac{2}{16\pi^2 \ep'} \left( \frac{11}{30} g_1^2 + \frac32 g_2^2 + \frac43 g_3^2 \right), \\
Z_{{\cal C}_{2}} & = 1 - \frac{2}{16\pi^2 \ep'} \left( \frac{23}{30} g_1^2 + \frac32 g_2^2 + \frac43 g_3^2 \right). \\
}
These are consistent with the results of Ref.~\cite{Munoz:1986kq}. In the next section, we determine the threshold corrections for the wave functions and the Wilson coefficients of the dimension-six baryon-number violating operators by matching the full and effective theories.

\section{ Threshold Corrections of the Dimension-Six Operators \label{sec:results}}

In the previous section, we have shown the radiative corrections to two-, three-, and four-point vertex functions in the SUSY $SU(5)$ GUTs and we have shown also the radiative corrections to the Wilson coefficients of the dimension-six operators in the EFT. Now, we determine the threshold corrections by matching the amplitudes in the EFT and those in the full theory.

First, let us discuss the threshold corrections to the two-point functions for matter superfields.
As we have seen in \cref{eq:fff}, the one-loop two-point functions are divided into two parts:
one is linear to $f(M_X^2)$ and the other is linear to $f(\mu_{\rm IR}^2)$.
The latter is the contribution from the MSSM gauge interactions, and the former is the contribution from the broken gauge interaction in $SU(5)$. On the other hand, the two-point functions in the EFT at the GUT scale have the form;
\eqs{
\Gamma^{\text{2-pt;eff}}_{\Phi} = (1-\lam_{\Phi}) \left( 1 + b_{\Phi}  f\left(\mu_{\rm IR}^2\right) \right) \Gamma^{\text{2-pt}}_{ \Phi \, 0}.}
Here, the chiral superfield
in the EFT is given by $(1-\lam_{\Phi}/2) \Phi$ ($\Phi$ is in the full theory). $\lam_\Phi$ is determined so as to match the two-point function in the EFT and that in the full theory:
\eqs{
\lam_{\Phi}(\mu) =   \frac{g_5^2}{16\pi^2} \Hat{\lambda}_{\Phi} f\left(M_X^2\right),
}
where 
$(\Hat{\lambda}_Q, \Hat{\lambda}_U, \Hat{\lambda}_D, \Hat{\lambda}_L, \Hat{\lambda}_E)=(3,4,2,3,6)$ is defined.

Next, we determine the threshold corrections for the baryon-number violating dimension-six operators. The two-point functions of the matter superfields in the full theory and the EFT are matched above, and we have determined the threshold corrections to the renormalizable kinetic terms.  For a matter superfield $\Phi$, the renormalizable kinetic term has the form $(1-\lam_\Phi) \Phi^\dag \Phi$ in the EFT. 
The finite corrections to the two-point functions in the EFT appear in the correction to the Wilson coefficients of higher-dimensional operators. The Wilson coefficients of higher-dimensional operators themselves also include the finite corrections. 
 Thus, we redefine the effective K\"ahler potentials ${\cal K}^\text{eff}_{I}$ $(I=1, \, 2)$ as the ones with threshold corrections up to the one-loop level as follows: 
\eqs{
{\cal K}^\text{eff}_{1} & = {\cal K}^\text{(1):eff}_{1} + \left(1-\lam_{1} - \frac12 (\lam_U + \lam_Q + \lam_D + \lam_L) \right){\cal K}^{(0)}_{1} \\
& ~~~~~ + \left( Z_{{\cal C}_{1}} {Z^{\text{EFT}}_U }^{1/2}{Z^{\text{EFT}}_Q }^{1/2}{Z^{\text{EFT}}_D }^{1/2}{Z^{\text{EFT}}_L }^{1/2} -1 \right) {\cal K}^{(0)}_{1}, \\
{\cal K}^\text{eff}_{2} & = {\cal K}^\text{(1):eff}_{2} +\left(1-\lam_{2} - \frac12 (\lam_E + \lam_U + 2 \lam_Q ) \right){\cal K}^{(0)}_{2} \\
& ~~~~~ + \left( Z_{{\cal C}_{2}} {Z^{\text{EFT}}_E }^{1/2}{Z^{\text{EFT}}_U }^{1/2}Z^{\text{EFT}}_Q  -1 \right) {\cal K}^{(0)}_{2}.
}
$\lambda_1$ and $\lambda_2$ are the threshold corrections to the Wilson coefficients for the baryon-number violating operators.

In the full theory (the SUSY $SU(5)$ GUTs), we have computed the effective K\"ahler potential for the dimension-six operators at the one-loop level,
\eqs{
{\cal K}_{1}^\text{full} = 
& ~ - \frac12\frac{1}{M_X^2+\Si(0)} {\cal C}_{V_2} {\cal C}_{V_1} {\cal O}^{(1)} + {\cal K}_{1}^{\text{(Box)}},  \\
{\cal K}_{2}^\text{full} = 
&  ~ - \frac12\frac{1}{M_X^2+\Si(0)} {\cal C}_{V_2} {\cal C}_{V_3}  {\cal O}^{(2)} + {\cal K}_{2}^{\text{(Box)}}. \\
}
The first terms include the vacuum polarization of the $X$ boson $\Si(0)$ and the one-loop effective couplings ${\cal C}_{V_1}, {\cal C}_{V_2}$, and ${\cal C}_{V_3}$ which are defined in \cref{eq:ver_corr}.

There are IR divergences in ${\cal K}_{I}^\text{full} $ and ${\cal K}_{I}^\text{eff} $ ($I=1, \, 2$), which are
represented by $\mu_{IR}$. The divergences are absorbed by the operators ${\cal O}^{(I)}$.\footnote{
Since the IR divergent terms from the box-like diagrams are proportional to $\ln M_X^2/\mu_{\rm IR}^2$, we divide this into $\ln M_X^2/\mu^2 + \ln \mu^2/\mu_{\rm IR}^2$ where $\mu$ denotes the renormalization scale, and then the IR divergent terms are absorbed by the operators.
}
Then, we divide the effective K\"ahler potentials into the coefficients ${\cal C}_{I}$ and the renormalized operator ${\cal O}^{(I)}_r$:
\begin{equation}
{\cal K}_{I}^\text{full}={\cal C}^{\text{full}}_{I} {\cal O}_r^{(I)},~ 
{\cal K}_{I}^\text{eff}={\cal C}^{\text{eff}}_{I} {\cal O}_r^{(I)}.
\end{equation}
The one-loop coefficients in the full theory are given by
\eqs{
\frac{{\cal C}^{(1):\text{full}}_{1}}{{\cal C}^{(0)}_{1}} & = \frac{M_X^2}{M_X^2+\Si(0)} - \frac{g_5^2}{16\pi^2} \left[ 6 + 6 \left(1-\ln\frac{M_X^2}{\mu^2} \right) - \frac{16}5 \ln\frac{M_X^2}{\mu^2}\right], \\
\frac{{\cal C}^{(1):\text{full}}_{2}}{{\cal C}^{(0)}_{2}} & =  \frac{M_X^2}{M_X^2+\Si(0)} - \frac{g_5^2}{16\pi^2} \left[ \frac{32}5 + 8 \left(1-\ln\frac{M_X^2}{\mu^2} \right) - \frac{18}5 \ln\frac{M_X^2}{\mu^2}\right],
}
where ${\cal C}^{(0)}_{1}$ and ${\cal C}^{(0)}_{2}$ are the tree-level ones: ${\cal C}^{(0)}_{1}={\cal C}^{(0)}_{2}=-g^2_5/M^2_X$.
In the EFT, the coefficients are 
\eqs{
\frac{{\cal C}^{(1):\text{eff}}_{1}}{{\cal C}^{(0)}_{1}} = & ~ 1 - \lam_{1}  - \frac{6 g_5^2}{16\pi^2} \left(1 - \ln\frac{M_X^2}{\mu^2} \right) - \frac{14}5 \frac{g_5^2}{16\pi^2} ,  \\
\frac{{\cal C}^{(1):\text{eff}}_{2}}{{\cal C}^{(0)}_{2}} = & ~ 1 - \lam_{2} - \frac{8 g_5^2}{16\pi^2} \left(1 - \ln\frac{M_X^2}{\mu^2} \right) - \frac{14}5 \frac{g_5^2}{16\pi^2} . 
}
We assume that the matching scale is $\mu= M_{\rm GUT} (\simeq M_X)$, where the unification $g_1= g_2= g_3= g_5$ is achieved. By comparing the amplitudes obtained in the full and effective theories, we determine the threshold corrections to the Wilson coefficients of dimension-six operators $\lam_{1}$ and  $\lam_{2}$ at the one-loop level:
\eqs{
\lam_{1} & = \frac{\Si(0)}{M_X^2+\Si(0)} + \frac{g_5^2}{16\pi^2} \frac{16}5 \left( 1 - \ln \frac{M_X^2}{\mu^2} \right), \\
\lam_{2} & = \frac{\Si(0)}{M_X^2+\Si(0)} + \frac{g_5^2}{16\pi^2} \frac{18}5 \left( 1 - \ln \frac{M_X^2}{\mu^2} \right).
}
We find that the corrections to the wave function for the matter field and the vertex of the $X$ boson are canceled with each other as expected from the Ward identity and that the threshold corrections come from the corrections to the vacuum polarization  and the box-like contributions.

Now, we give numerical results of the short-range renormalization factor including threshold corrections in the minimal SUSY $SU(5)$ GUT and its vector-like extension. In the minimal SUSY $SU(5)$ GUT, the $X$ multiplet, the color-triplet Higgs multiplets, and the adjoint Higgs multiplet acquire heavy mass through the VEV of the adjoint Higgs multiplet. First, we set the masses of the GUT particles to be degenerate in mass $2.0\times10^{16}~{\rm GeV}$ since they are model-dependent parameters. The dependence of the threshold correction on the GUT scale mass spectrum is shown later. 
The threshold corrections in the minimal SUSY $SU(5)$ GUT are divided into two parts: the one comes from the vacuum polarization of the $X$ boson as
\eqs{
\left.\lam_1\right|_{\textrm{vac.}} = \left.\lam_2\right|_{\textrm{vac.}} = \frac{\Si(0)}{M_X^2+\Si(0)} = -3.68 \times 10^{-2},
\label{eq:lamvac_MSGUT}
}
another one comes from the box-type diagram:
\eqs{
\left.\lam_1\right|_{\textrm{vert.}} & = \frac{g_5^2}{16\pi^2} \frac{16}5 \left(1 - \ln \frac{M_X^2}{\mu^2} \right) = 1.03 \times 10^{-2}, \\
\left.\lam_2\right|_{\textrm{vert.}} & = \frac{g_5^2}{16\pi^2} \frac{18}5 \left(1 - \ln \frac{M_X^2}{\mu^2} \right) = 1.15 \times10^{-2}.
}
Then, by combining these contribution we obtain the numerical values of threshold corrections as
\eqs{
\lam_{1}(M_{\rm GUT}) = -2.66\times10^{-2}, ~~~~~
\lam_{2}(M_{\rm GUT}) = -2.53\times10^{-2},
}
where we assume that all sparticle masses are set to be $M_{\rm SUSY} = 1~{\rm TeV}$. We set the renormalization scale at which we match the amplitudes in the full theory and the EFT to $M_{\rm GUT} = 2.0 \times 10^{16}~{\rm GeV}$. 

The short-range renormalization factors of Wilson coefficients of the dimension-six operators which include two-loop RGEs and threshold corrections are defined as:
\eqs{
A_S^{(I)} \equiv (1-\lam_{I}) \frac{{\cal C}_{I}(M_{\rm SUSY})}{{\cal C}_{I}(M_{\rm GUT})},
}
where ${\cal C}_I(\mu)$ are the Wilson coefficients of the dimension-six operators at renormalization scale $\mu$, which do not include threshold corrections at GUT scale.
These numerical factors are obtained by using 
the RGEs at the two-loop level\footnote{
The RGEs for the gauge and Yukawa coupling constants and the Wilson coefficients for the baryon-number violating dimension-six operators  are summarized in \cref{app:RGE}.}
\eqs{
A_S^{(1)} = 2.025, ~~~~~~
A_S^{(2)} = 2.118.
}

We have also evaluated the short-range renormalization factor to the partial decay rate ($p \to e^+ + \pi^0$). We define the ratio of the short-range renormalization factor with and without the threshold correction to the Wilson coefficients of the dimension-six operators as
\eqs{
R \equiv \frac{\left. A_S^{(1)2}+(1+|V_{ud}|^2)^2A_S^{(2)2}\right|_{\rm w}}{\left. A_S^{(1)2}+(1+|V_{ud}|^2)^2A_S^{(2)2}\right|_{\rm w/o}},
}
where the denominator and numerator correspond to the short-range enhancement factor of the nucleon decay rate without and with threshold corrections, respectively. $V_{ud}$ denotes $(1,1)$ component of the CKM matrix. We obtain $R=1.052$ in the minimal SUSY $SU(5)$ GUT, that is, there is about 5\% enhancement compared with the short-range renormalization factor without threshold corrections.

We note that the mass relation between $M_\Si$ and $M_{\Si_{24}}$ is $M_\Si = 5 M_{\Si_{24}}$ in the minimal SUSY $SU(5)$ GUT. When we adopt this mass relation and we assume the masses of the GUT particles are set to be $M_X = M_\Si = 2.0 \times 10^{16}~{\rm GeV}$, we have
\eqs{
A_S^{(1)} = 2.014, ~~~~~~
A_S^{(2)} = 2.107,
}
and then, we obtain $R = 1.041$.

\begin{figure}[t]
\centering
\includegraphics[width=6cm,angle=0]{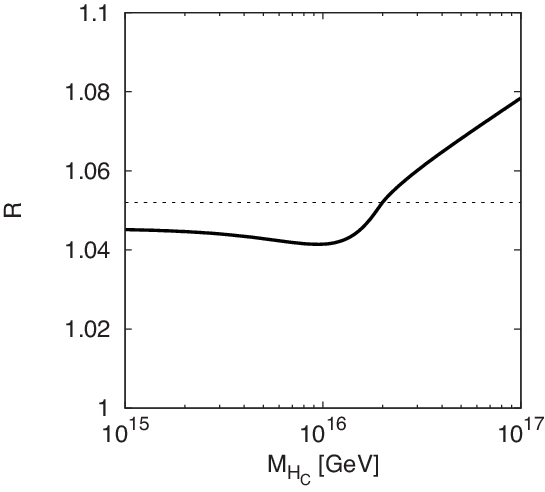} ~~~~~
\includegraphics[width=6cm,angle=0]{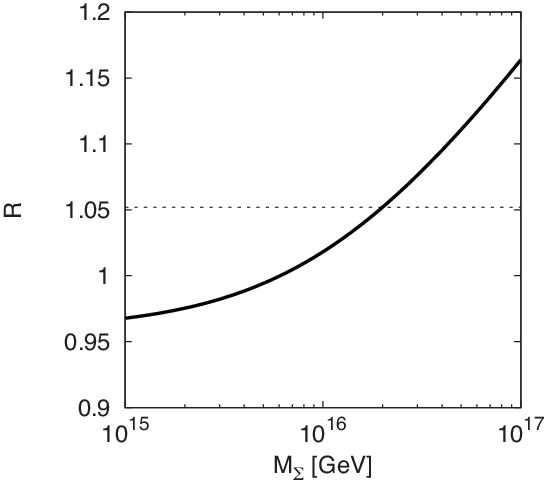}
\caption{GUT-scale particle mass dependence on renormalization factor of proton decay. Left panel shows $M_{H_C}$ dependence for $M_\Si=M_X=2\times10^{16}~{\rm GeV}$. Right panel shows $M_\Si$ dependence for  $M_{H_C}=M_X=2\times10^{16}~{\rm GeV}$. Dotted line shows the degenerate mass case in each panels.}
\label{fig:massdep}
\end{figure}

In \cref{fig:massdep}, we describe the heavy mass dependence on the ratio of the short-range renormalization factor in the minimal SUSY $SU(5)$ GUT. Here, we set the mass of the component fields of the adjoint Higgs multiplet to be degenerate in $M_\Si$, that is, we set $M_{\Si_{24}} = M_\Si$, for simplicity. The left panel of \cref{fig:massdep} shows the color-triplet Higgs mass ($M_{H_C}$) dependence of the ratio with the fixed adjoint Higgs mass $M_\Si = 2.0\times10^{16}~{\rm GeV}$. The right panel of \cref{fig:massdep} shows the adjoint Higgs mass ($M_\Si$) dependence of the ratio with the fixed color-triplet Higgs mass $M_{H_C} = 2.0\times10^{16}~{\rm GeV}$. 
Since, in a large $M_{H_C}$ region, the vacuum polarization behaves as $\Si \sim \frac12 M_X^2 (\frac12 - \ln M_{H_C}^2/\mu^2)$, the decay rate of proton is slightly enhanced in this region.

\begin{figure}[t]
\centering
\includegraphics[width=8cm,angle=0]{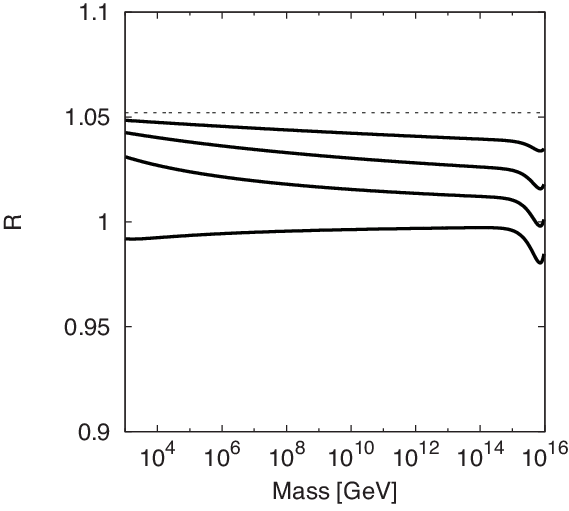}
\caption{Ratio of short-range renormalization effects with and without threshold effect in the minimal SUSY $SU(5)$ GUT with light vector-like matters. We take $n_{\bold 5} = 1, \cdots, 4$ in solid lines from top to bottom. The case of the minimal SUSY $SU(5)$ with no light vector-like matter is shown in dotted line.}
\label{fig:ratio}
\end{figure}

In the SUSY $SU(5)$ GUT with light vector-like matter scenario, the threshold corrections to the Wilson coefficients of the dimension-six operators are enhanced since the unified gauge coupling becomes large. This large unified coupling leads to the large renormalization effect to the Wilson coefficients of the dimension-six operators.

In \cref{fig:ratio}, we show the ratio of the short-range renormalization factors in the vector-like matter scenario. The horizontal line and the vertical line present the mass scale of the vector-like matters and the ratio of the short-range renormalization effect, respectively. The solid lines correspond to the case that the number of $\bold{5} + \ovl{\bold{5}}$ vector-like matters is set to be $n_{\bold 5} = 1, \cdots, 4$ from top to bottom without $\bold{10} + \ovl{\bold{10}}$ vector-like matter.
In this estimation, we assume the masses of the heavy multiplets and the GUT scale are set to be $2.0 \times 10^{16}$~GeV. If the mass (number) of the vector-like superfields is sufficiently light (large), the unified gauge coupling at the GUT scale becomes larger.
However, the additional contribution from the vector-like matters cancels with the gauge contributions.
In fact, the vacuum polarization $\Si_X(0)$ from the vector-like matters is proportional to 
\eqs{
\left. \Si_X(0)\right|_{\text{vector-like}} \propto \frac{g_5^2(N_{\mathbf{5}}+N_{\ovl{\mathbf{5}}})}{16\pi^2} M_X^2 \left(1-\ln\frac{M_X}{\mu} \right).
}
Here, we neglect the vector-like mass dependence since we are interested in the case of the sufficiently small vector-like masses.
Therefore, the additional positive contribution from the vector-like matters cancels with the negative contribution $\left. \lam_i \right|_{\rm vec.} ~ (i = 1, 2)$ in \cref{eq:lamvac_MSGUT} when we set $M_X = \mu = 2.0 \times 10^{16}~{\rm GeV}$.

\begin{figure}[t]
\centering
\includegraphics[width=10cm,angle=0]{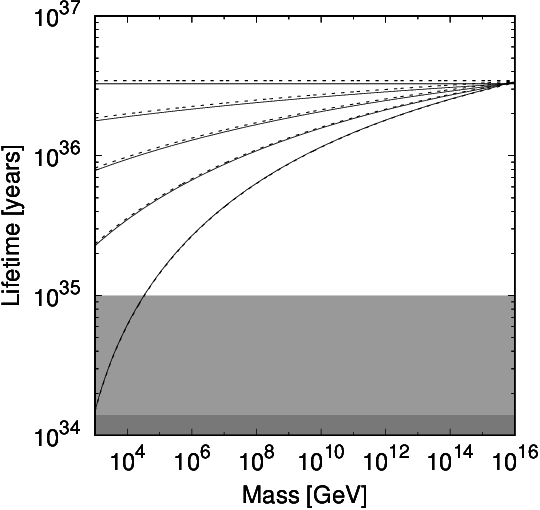}
\caption{Partial proton lifetime ($p \to \pi^0 + e^+$) in vector-like extension scenario. In solid (dotted) lines, we take $n_{\bold 5} = 0,1,\cdots, 4$ with (without) threshold corrections at GUT scale. Deep gray (gray) region corresponds to experimental excluded region by Super-Kamiokande (the future sensitivity by the Hyper-Kamiokande).}
\label{fig:lifetime}
\end{figure}

In \cref{fig:lifetime}, we show the partial proton lifetime ($p \to \pi^0 + e^+$) in the minimal SUSY $SU(5)$ and its vector-like extension. In this evaluation, we assume the masses of the GUT spectrum are set to be the same mass $(2.0 \times 10^{16}~{\rm GeV})$, especially the $X$-boson mass is set to be $M_X = 2.0 \times 10^{16}$~GeV. We use the two-loop RGEs of the Wilson coefficients of the dimension-six operators as short-distance \cite{Daniel:1983ip,Hisano:2013ege} and as long-distance \cite{Nihei:1994tx}. We also use the hadron matrix elements evaluated with the lattice calculation \cite{Aoki:2013yxa}. The deep gray region is corresponding to the present lower bound on this decay mode by the Super-Kamiokande ($\tau (p \to \pi^0 + e^+) > 1.4 \times 10^{34}$ years). The gray region, on the other hand, corresponds to the future sensitivity on this decay mode by the Hyper-Kamiokande ($\tau (p \to \pi^0 + e^+) > 1.0 \times 10^{35}$ years). Due to the extra fields, the lifetime is suppressed since the unified coupling becomes large at GUT scale.

\section{Conclusion and Discussion  \label{sec:conclusion}}

In this study, we have derived the threshold corrections to the Wilson coefficients which cause proton decay ($p\to\pi^0+e^+$) at the GUT scale in SUSY $SU(5)$ GUTs. We find that the threshold correction makes the proton decay rate enhanced about 5\% in the minimal SUSY $SU(5)$ GUT. Furthermore, we also have investigated the threshold effect on the partial proton decay rate in the extended SUSY $SU(5)$ GUT with additional vector-like pairs, motivated by the achievement of the 126~GeV Higgs boson. 
In these models, we find that the threshold corrections give tiny effects in spite of the large unified gauge coupling. 
This is due to the cancellation between contributions from additional vector-like matters and gauge multiplets.

In our study, we neglect the threshold corrections induced by the Yukawa interactions, because the Yukawa interactions involving light quarks and leptons are negligibly small at the GUT scale.
Similarly, we do not estimate the threshold correction at the scale where superparticles are decoupled.
In this work, we have concentrated on the effect of vector-like matters at the GUT scale.
In order to complete the evaluation of two-loop level corrections, we should include the one-loop threshold correction at the SUSY scale.
We will calculate these corrections on another occasion.

There exists the additional loop suppression in the next-to-next-to leading order (NNLO) calculations such as three-loop RGEs and two-loop threshold corrections.
The loop factor at the GUT scale, that is $g_5^2(M_{\rm GUT})/16\pi^2$, becomes $3.3 \times 10^{-3}$ to $1.5 \times 10^{-2}$ corresponding to the number of vector-like matters being $N_5 = 0$ to $N_5 = 4$.
Thus, the NNLO calculations should be much smaller than the uncertainty of the matrix elements derived by using lattice QCD simulation as discussed below.

The matrix elements relevant to nucleon decay have been evaluated with the lattice QCD and they have 30\% uncertainty at present \cite{Aoki:2013yxa}.
In this work, we have revealed that the corrections in the minimal SUSY $SU(5)$ GUT and its vector-like extensions are small in comparison with the uncertainty of the matrix elements.
We expect that the uncertainty would be reduced in the future.

Finally, we note the application of our work to the other SUSY GUTs. We only have investigated the threshold effects in the minimal SUSY $SU(5)$ GUT and the extra vector-like matter extensions in this paper. 
When, however, we apply our formulae for the extension of the SUSY $SU(5)$ GUTs, for instance the missing-partner model \cite{Hisano:1994fn}, we only have to evaluate additional contributions to the vacuum polarization for the $X$ boson. That is remaining as one of our future work.

\section*{Acknowledgements}
This work is supported by Grant-in-Aid for Scientific research from the Ministry of Education, Science, Sports, and Culture (MEXT), Japan, No. 24340047 (for J.H.) and No. 23104011 (for J.H. and Y.O.). The work of J.H. is also supported by World Premier International Research Center Initiative (WPI Initiative), MEXT, Japan.

\newpage
\appendix
\renewcommand{\theequation}{\Alph{section}.\arabic{equation}}
\setcounter{equation}{0}
\section{Decomposition of $SU(5)$ Interactions \label{ap:gauge int}}
\subsection{Interactions of Vector Superfields}
In super-Yang-Mills theories, the renormalizable Lagrangian is written as
\eqs{
{\cal L}_{\rm SYM} = \frac{1}{8g^2} \tr \int d^2\theta {\cal W}^{\alpha} {\cal W}_\alpha  + {\rm h.c.},
}
where the field strength chiral superfield is given in \cref{eq:field_strength}. The Lagrangian is expanded in the vector superfield $V$ as 
\eqs{
\frac{1}{8g^2} \tr \int d^2\theta {\cal W}^{\alpha} {\cal W}_\alpha 
= & - \frac{1}{8} \tr \int d^4\theta \left[ - V \coD^\alpha \ovl \coD^2 \coD_\alpha V + 2 g V \coD^\alpha \ovl \coD^2 [ V, \coD_\alpha V]   \right. \\ 
& \left. + g^2 [ V, \coD^\alpha V] \ovl \coD^2 [ V, \coD_\alpha V] + \frac{4g^2}{3} \coD_\alpha V \ovl \coD^2 [ V, [ V, \coD_\alpha V]] + \cdots \right]. 
\label{eq:kinetic_gf}
} 

The decomposition of the $SU(5)$ vector superfield ${\cal V}_5$ is given by \cref{eq:decomp_V}.
As mentioned in text, we denote $SU(3)_C, SU(2)_L$, and $U(1)_Y$ vector superfields in the MSSM with $G, W$, and $B$.
The kinetic terms of the vector superfields in the $SU(5)$ GUTs are given into the following form;
\eqs{
{\cal L}_{VV}
& =  2 \tr \int d^4\theta ~ G \square P_T G + 2 \tr \int d^4\theta ~ W \square P_T W +  \int d^4\theta ~ B \square P_T B + 2 \int d^4\theta ~ X^{\dag} \square P_T X,
}
where $X$ denotes the massive vector superfield associated with the broken $SU(5)$ generators. Here, $P_T(\equiv\coD^\alpha \ovl \coD^2\coD_\alpha/(8{\square}))$ is the projection operator to the transverse mode ($P_T^2=P_T$).

From the second term of \cref{eq:kinetic_gf}, 
the three-point interaction terms between $X$ and MSSM vector superfields are obtained as 
\eqs{
{\cal L}_{X\text{-3 pt}} = \int d^4\theta \left[ \del^s_r (T^a)_\alpha^\beta ({\cal K}_{XG}^a)^{r\alpha}_{s\beta} + (t^a)^s_r \del_\alpha^\beta ({\cal K}_{XW}^a)^{r\alpha}_{s\beta} + \frac{5}{2\sqrt{15}} \del^s_r \del_\alpha^\beta ({\cal K}_{XB})^{r\alpha}_{s\beta} \right],
}
where 
\eqs{
({\cal K}_{XV}^a)^{r\alpha}_{s\beta} \equiv \frac{g_5}{4} & \left[ X^r_\beta (\ovl \coD^2 \coD X^\dag)_s^\alpha \coD V^a 
+ V^a (\ovl \coD^2 \coD X)^r_\beta (\coD X^\dag)_s^\alpha 
+ {X^\dag}_s^\alpha (\ovl \coD^2 \coD V^a) \coD X^r_\beta \right. \\
& \left. - X^r_\beta (\ovl \coD^2 \coD V^a)(\coD X^\dag)_s^\alpha 
- V^a (\ovl \coD^2 \coD X^\dag)_s^\alpha \coD X^r_\beta
- {X^\dag}_s^\alpha (\ovl \coD^2 \coD X)^r_\beta \coD V^a \right]. \\
}
Here, spinor indices are contracted like ${{}^\alpha}_\alpha$ or ${{}_{\dot\alpha}}^{\dot\alpha}$.
The four-point self interaction of $X$ is given as 
\eqs{
{\cal L}_{X\text{-4 pt}} = & - \frac{g_5^2}{48} \int d^4 \theta\left[ (\ovl \coD^2 \coD {X^\dag}^\alpha_r) ( X^r_\beta {X^\dag}^\beta_s (\coD X)^s_\alpha - 2 X^r_\beta (\coD X^{\dag})^\beta_s X^s_\alpha + (\coD X)^r_\beta {X^{\dag}}^\beta_s X^s_\alpha) \right. \\
& + \left. (\ovl \coD^2 \coD X^r_\alpha) ({X^{\dag}}^\alpha_s X^s_\beta (\coD X^{\dag})^\beta_r -2 {X^{\dag}}^\alpha_s (\coD X)^s_\beta {X^{\dag}}^\beta_r + (\coD X^{\dag})^\alpha_s X^s_\beta {X^{\dag}}^\beta_r) \right].
}

\subsection{Vector-Ghost Interactions}
The Lagrangian for the massless Fadeev-Popov ghost chiral superfields, which are denoted by $b$ and $c$, are given as
\eqs{
{\cal L}_{\rm FP} = 2~\tr \int d^4\theta (b+b^\dag) {\cal L}_{g V} \left[ (c+c^\dag) + \coth({\cal L}_{g V})(c-c^\dag) \right],
\label{eq:ghost}
}
where ${\cal L}_A B$ is the Lie derivative (${\cal L}_A B \equiv [A,B]$). Therefore, the kinetic terms for ghost fields in the $SU(5)$  GUTs are obtained as
\eqs{
{\cal L}_{\text{ghost}}
= & 2 \int d^4\theta \left[ \tr(b_3^\dag c_3 - b_3 c_3^\dag) + \tr(b_2^\dag c_2 - b_2 c_2^\dag) \right] 
+  \int d^4\theta (b_1^\dag c_1 - b_1 c_1^\dag) \\
& + \int d^4\theta \left[ (b_X^\dag c_X - b_X c_X^\dag) + (b_{X^\dag}^\dag c_{X^\dag} - b_{X^\dag} c_{X^\dag}^\dag) \right], \\
}
where the ghost multiplets are decomposed in a similar way to the gauge multiplets as
\eqs{
b = \left( 
\begin{array}{cc}
b_3 - \frac{2}{\sqrt{30}} b_1 &  \frac1{\sqrt{2}} b_X \\
 \frac1{\sqrt{2}} b_{X^\dag} & b_2 + \frac{3}{\sqrt{30}} b_1
\end{array}
\right), ~~~~~
c = \left( 
\begin{array}{cc}
c_3 - \frac{2}{\sqrt{30}} c_1 &  \frac1{\sqrt{2}} c_X \\
\frac1{\sqrt{2}} c_{X^\dag} & c_2 + \frac{3}{\sqrt{30}} c_1
\end{array}
\right).
}

After spontaneously breaking of the GUT group by the adjoint Higgs chiral superfield, there exist kinetic mixing terms between $X$ and the Nambu-Goldstone chiral superfields $\Si_{(3,2)}$ and $\Si_{(3^\ast,2)}$. By using the supersymmetric $R_\xi$-gauge \cite{Ovrut:1981wa}, we remove the kinetic mixing terms, and we find the mass terms for the ghost chiral superfields \cite{Ovrut:1981wa} as:
\eqs{
{\cal L}_{\text{ghost mass}} = \int d^4\theta \left[  (b_X + b_X^\dag) \frac{M_X^2}{\xi \square} (c_X-c_X^\dag) +  (b_{X^\dag} + b_{X^\dag}^\dag) \frac{M_X^2}{\xi \square} (c_{X^\dag}-c_{X^\dag}^\dag)\right].
}
We note that the terms such as $b_X c_X$ and $b_X^\dag c_X^\dag$ vanish by the superspace integral since these are chiral (or antichiral) superfields. Then, the propagator for massive ghost superfields is modified as
\eqs{
\Delta_{bc} = \frac{i}{k^2} \del^4(\theta_1-\theta_2) \to \frac{i}{k^2} \frac{1}{1- \frac{M_X^2}{\xi k^2}}\del^4(\theta_1-\theta_2).
}

In the evaluation of the self energy of $X$, we need interaction terms for $X$ and the massive ghosts. In general, three-point and four-point interaction terms of ghost superfields and vector superfields are obtained from \cref{eq:ghost} as follows,
\eqs{
{\cal L}_{bVc} = \tr \int d^4\theta \left \{ 2g ~ (b+b^\dag) \left[ V, (c+c^\dag) \right] + \frac{2g^2}{3} (b+b^\dag)[V,[V,(c-c^\dag)]] + {\cal O}(V^3)\right\}.
}
Then, the interaction terms between $X$ and the ghosts are given by:
\eqs{
{\cal L}_{bXc} = \int d^4 \theta \left[ \del^s_r (T^a)_\alpha^\beta ({\cal K}_{bcG}^a)^{r\alpha}_{s\beta} - (t^a)^s_r \del_\alpha^\beta ({\cal K}_{bcW}^a)^{r\alpha}_{s\beta} - \frac{5}{2\sqrt{15}} \del^s_r \del_\alpha^\beta ({\cal K}_{bcB})^{r\alpha}_{s\beta} \right],
}
and
\eqs{
{\cal L}_{bX^2c} = & - \frac{g_5^2}{6} \int d^4\theta (\del^{\beta\del}_{\alpha\gamma} \del^{tr}_{su} + \del^{\del\beta}_{\alpha\gamma} \del^{rt}_{su}) X^{\dag\alpha}_r X^s_\beta \\
& \times \left[ (b_{X^\dag})^\gamma_t (c^\dag_{X^\dag})^u_\del - (b_X^\dag)^\gamma_t (c_{X})^u_\del - (b^\dag_{X^\dag})^u_\del (c_{X^\dag})^\gamma_t  + (b_X)^u_\del (c_X^\dag)^\gamma_t \right] .
}
Here, we define $\del^{\beta\del}_{\alpha\gamma} \equiv \del^{\beta}_{\alpha} \del^{\del}_{\gamma}$ and $\del^{tr}_{su}\equiv \del^{t}_{s}\del^{r}_{u}$. 
In the three-point interactions, we define the term $({\cal K}_{bcV}^a)^{r\alpha}_{s\beta}$ as:
\eqs{
({\cal K}_{bcV}^a)^{r\alpha}_{s\beta} & \equiv 
((b_X+b_{X^\dag})^r_\beta {X^\dag}^\alpha_s - X^r_\beta (b_{X^\dag}+b_X^\dag)^\alpha_s) (c_V+c_V^\dag)^a \\
& + (b_V+b_V^\dag)^a (X^r_\beta (c_{X^\dag}+c_X^\dag)^\alpha_s - (c_X + c_{X^\dag}^\dag)^r_\beta {X^\dag}^\alpha_s).
}

\subsection{Gauge Interactions of Matter Superfields}
Now, we summarize the gauge interactions of the matter and Higgs multiplets in SUSY $SU(5)$ GUTs.
The renormalizable K\"ahler potential in the $SU(5)$ GUTs is given as:
\eqs{
{\cal K} = & ~ \Phi^{\dag A} (e^{-2g_5{\cal V}_5})_A^B \Phi_B 
+ \Psi^\dag_{AB} (e^{2g_5{\cal V}_5})^A_C (e^{2g_5{\cal V}_5})_D^B \Psi^{CD} \\ 
& + 2 \Si^{\dag A}_B (e^{-2g_5{\cal V}_5})^C_A (e^{2g_5{\cal V}_5})_D^B \Si^D_C 
+ H^{\dag A}_{\ovl{\bold 5}} (e^{-2g_5{\cal V}_5})_A^B H_{\ovl{\bold 5} B} 
+ H^\dag_{\bold{5} A} (e^{2g_5{\cal V}_5})^A_B H^B_{\bold 5}.
}

The three-point gauge interaction of the $\ovl{\bold{ 5}}$ representation matter field $\Phi$ is given as 
\eqs{
{\cal K}_{\Phi^\dag V \Phi} = & - g_5 D^{C\dag} \left( 2 G - \frac {2}{\sqrt{15}} B \right)D^C
+ g_5 L^\dag \left (2W - \frac {3}{\sqrt{15}} B\right) L \\
& - \sqrt 2 g_5 \left[ D^{C\dag } (X\cdot L) + {\rm h.c.} \right]. \\
}
For the four-point vertices, we only use the interactions which include only one $X$,
\eqs{
{\cal K}_{\Phi^\dag V^2 \Phi} \ni & ~~ 
\sqrt 2 g_5^2 \left(D^{C\dag} G (X\cdot L)
+ \frac{1}{\sqrt{60}} D^{C\dag} B (X\cdot L)
+ D^{C\dag }(WX\cdot L)\right)
+ {\rm h.c.} \, . \\
}
Here, $(A \cdot B)\equiv \ep_{rs}A^rB^s$.  We also obtain the relevant
gauge interactions from the $\bold{ 10 }$ representation matter field
$\Psi$,
\eqs{
  {\cal K}_{\Psi^\dag V \Psi} = & - g_5 U^{C\dag} \left(2 G+\frac {4}{\sqrt{15}} B\right) U^C 
 + g_5  Q^\dag \left(2 G+ 2W + \frac{1}{\sqrt{15}} B\right) Q \\
& + g_5 \frac {6}{\sqrt{15}} E^{C\dag} B E^C 
 + \sqrt2 g_5 \left[ [Q^\dag X U^{C}] - (Q^\dag\cdot X^{\dag}) E^C + {\rm h.c.} \right], \\
}
\eqs{
{\cal K}_{\Psi^\dag V^2 \Psi} \ni & ~~ \sqrt 2 g_5^2 \left[ 
[(GQ^\dag)XU^C] - [Q^\dag X(GU^C)] -  \frac{3}{\sqrt{60}} B[Q^\dag XU^C] 
+ [(WQ^\dag)XU^C] \right.\\
&
\left.
+ E^{C\dag} \left((X\cdot GQ)
+ (X\cdot WQ)
+ \frac{7}{\sqrt {60}} (X\cdot BQ) \right)
\right]  + {\rm h.c.}\, , \\
&
}
where 
$[ABC]\equiv\ep^{\alpha\beta\gamma}A_{\alpha} B_{\beta} C_{\gamma}$ or $\ep_{\alpha\beta\gamma}A^{\alpha} B^{\beta} C^{\gamma}$.

There are also the three- and four-point interactions with Higgs multiplets of $X$. One of those comes from the interaction of the anti-fundamental Higgs superfield $\ovl H = (H_{\ovl C}, H_d)$,
\eqs{
{\cal K}_{{{\ovl H}^\dag} X \ovl H} = & - \sqrt 2 g_5 \left[ H^\dag_{\ovl C} (X\cdot H_d) + {\rm h.c.} \right] \\
& + g_5^2 \left[
\sqrt 2 \left( H^\dag_{\ovl C}G(X\cdot H_d)
+ \frac{1}{2\sqrt{15}}H^\dag_{\ovl C}B(X\cdot H_d)
+ H^\dag_{\ovl C}(WX\cdot H_d)\right)
+ {\rm h.c.} \right. \\
& + \left. H_{\ovl C}^{\dag\alpha} X_\alpha^r  X^{\dag\beta}_r H_{\ovl C\beta}
+  (X^\dag \cdot H^\dag_{d}) (X \cdot H_d) \right].
}
Another one comes from the fundamental Higgs superfield $H = (H_C, H_u)$,
\eqs{
{\cal K}_{H^\dag X H} = & \sqrt 2 g_5 \left[ H^\dag_{u} X H_C + {\rm h.c.} \right] \\
& + g_5^2 \left[ 
\sqrt 2 \left(H^\dag_{u} XGH_C
+ \frac{1}{2\sqrt{15}} H^\dag_{u} XBH_C
+ H^\dag_{u} W X H_C\right)
 + {\rm h.c.} \right. \\
& + \left. H^\dag_{C\alpha} X_r^{\dag\alpha} X^r_\beta H_C^\beta
+ H^\dag_{ur} X^r_\alpha X^{\dag\alpha}_s H_u^s \right].
}
The adjoint Higgs superfield is decomposed as 
\eqs{
\Si = \left( 
\begin{array}{cc}
\dps \Si_8 - \frac{2}{\sqrt{60}} \Si_{24} & \dps \frac{1}{\sqrt{2}} \Si_{(3,2)} \\
\dps \frac{1}{\sqrt{2}} \Si_{(3^\ast,2)} & \dps \Si_3 + \frac{3}{\sqrt{60}} \Si_{24}
\end{array}
\right).
}
In our calculation, we need the interaction terms with the adjoint Higgs superfield of $X$,
\eqs{
{\cal K}_{\Si^\dag X \Si} = & - 2 g_5 \left[ \Si^\dag_{(3^\ast,2)} X \Si_8 - \Si^\dag_8 X \Si_{(3,2)} \right]
 - 2 g_5 \left[ \Si^\dag_3 X\Si_{(3,2)} - \Si^\dag_{(3^\ast,2)}  X \Si_3 \right] \\
& - \frac{5}{\sqrt{15}} g_5 \left[ \Si^\dag_{(3^\ast,2)}  X \Si_{24} - \Si^\dag_{24} X\Si_{(3,2)} \right] + {\rm h.c.}, \\
}
\eqs{
{\cal K}_{\Si^\dag X^\dag X \Si} = & 
g_5^2 \left\{
2X^\dag ( \Si^\dag_8 \Si_8 + \Si_8 \Si^\dag_8)X
+2 X^\dag ( \Si^\dag_3 \Si_3 + \Si_3  \Si^\dag_3) X
 + \frac{5}{3} X^\dag \Si^\dag_{24} \Si_{24}  X \right.
\\
& + \frac{10}{\sqrt{15}} (\Si^\dag_{24} X^{\dag} X \Si_8 - \Si^\dag_{24} X^{\dag} X \Si_3 + {\rm h.c.}) \\
& - 2 \left( 2 \Si^\dag_8 X^{\dag} X \Si_3 + (\Si^\dag_{(3^\ast,2)})^\alpha_r X^s_\alpha X^r_\beta (\Si_{(3,2)})^\beta_s +{\rm h.c.}  \right) \\
& \left. + (\del^{ru}_{st} \del^{\alpha\gamma}_{\del\beta} + \del^{ru}_{ts} \del^{\alpha\gamma}_{\beta\del}) X^t_\gamma X^{\dag \del}_u \left( (\Si^\dag_{(3,2)})_\alpha^s  (\Si_{(3,2)})^\beta_r + (\Si^\dag_{(3^\ast,2)})^\beta_r (\Si_{(3^\ast,2)})_\alpha^s\right) \right\}.
 }
After symmetry breaking of GUT, there exist the three-point interaction terms between MSSM vector superfields, Nambu-Goldstone multiplet, and $X$ with VEV $v_\Si$ of the adjoint Higgs multiplet.
\eqs{
{\cal K}_{v_\Si X} & = - 10 g_5^2 v_\Si \left\{ 
\Si_{(3,2)} \left[ GX - W X + \frac{5}{\sqrt{60}} B X \right] \right. \\
& ~~~ \left. + \Si_{(3^\ast,2)} \left[ GX^\dag - W X^\dag + \frac{5}{\sqrt{60}} B X^\dag \right] 
\right\} + \text{ h.c.}.
}

\section{Radiative Corrections at One-loop \label{app:explicit_loops}}
In this appendix, we give the explicit formulae of the loop integrals in terms of supergraphs. All the external momenta of the chiral (antichiral) superfields are set to be $p$, and the masses of the MSSM vector superfields are set to be $\mu_{\rm IR}$ in order to regularize the IR divergence. For simplicity, we set all coupling constants to be 1 through this appendix. For the corrections to the three-point vertex functions and the box-like corrections, the loop integrals in text are the coefficients of K\"ahler potentials in the limit that the external momenta $p^2$ vanishes.

\subsubsection*{Radiative Corrections to Two-Point Functions for Matter Superfields}
The correction to the self energy of the chiral and antichiral matter superfields in the first generation is induced by the gauge interactions. The one-loop contribution is given as 
\eqs{
i\Gamma_\Phi & = i^2 \int d^4 \theta_1 d^4 \theta_2 \int \frac{d^D l}{ (2\pi)^D} \frac{-i}{2(l^2-M^2)}\frac{i}{(l+p)^2} \frac{1}{16}( \coD^2_2 \del_{21} \ola{\ovl \coD^2_1}) \del_{12} \Phi(p,\theta_1) \Phi^\dag(p,\theta_2) \\
& = - \frac{1}{2} \int \frac{d^D l}{ (2\pi)^D}\frac{1}{l^2-M^2}\frac{1}{(l+p)^2} \int d^4\theta \Phi^\dag(p,\theta)\Phi(p,\theta),
}
where $p$ is external momentum and $M$ is the mass for the internal vector superfield. $\del_{ij}$ denotes the $\delta$-function for the Grassmann valuable, $\del_{ij} \equiv (\del_i-\del_j)^2(\ovl\del_i-\ovl\del_j)^2$. The renormalized one-loop two-point function of matter superfields in the $SU(5)$ GUTs are given as 
\eqs{
\Gamma_{\Phi} & = - \frac{g_5^2}{8\pi^2} \left[ (c^{\Phi}_5 - \sum^3_{n=1} c^{\Phi}_n) f(M_X^2) + \sum^3_{n=1} c^{\Phi}_n f(\mu_{\rm IR}^2) \right] \int d^4 \theta \Phi^\dag \Phi,
}
where function $f(M^2)$ is defined in \cref{eq:fcs_wr}.  $c^{\Phi}_5$ and $c^{\Phi}_n$ ($n=3,\,2,\,1$) are the quadratic Casimir defined in text. In the MSSM, we also obtain
\eqs{
\Gamma_{\Phi}^{\rm EFT} & = - \frac{g_5^2}{8\pi^2} \sum^3_{n=1} c^{\Phi}_n f(\mu_{\rm IR}^2) \int d^4 \theta \Phi^\dag \Phi. \\
}

\subsubsection*{Radiative Corrections to Two-Point Function for Vector Superfield}

Three diagrams in \cref{fig:quad} contribute to the radiative corrections to two-point functions from the (massive) chiral superfields. The corrections from the diagram (a) in \cref{fig:quad} are 
\eqs{
i\Gamma_{XX}^{\text{(a)}} = - i^2 \int d^4 \theta X^{\dag \alpha}_r(-p,\theta) \int \frac{d^D l}{(2\pi)^D} \frac{l^2 - \frac12 l_\mu \si^\mu_{\alpha\dot\alpha} \coD^\alpha \ovl \coD^{\dot\alpha} + \frac{1}{16} \coD^2 \ovl \coD^2}{(l^2-M_1^2)[(l+p)^2-M_2^2]} X_\alpha^r(p,\theta),
}
where $M_1$ and $M_2$ are the masses of the chiral superfields in the loop diagram.
After picking the transverse mode and regularizing the UV divergence, we obtain the finite correction to the two-point function as follows:
\eqs{
\Gamma_{XX}^{\text{(a)}} = \frac{1}{16\pi^2} B(p^2,M_1^2,M_2^2) \int d^4 \theta X^{\dag \alpha}_r(-p,\theta) P_T X_\alpha^r(p,\theta)+{\text{(longitudinal mode)}},
}
where the loop function is defined in \cref{eq:loop_func}.
The massive chiral superfields also have the non-zero contribution from the diagram (b) in \cref{fig:quad},
\eqs{
\Gamma_{XX}^{\text{(b)}} = - \frac{M^2}{16\pi^2} \left( 1 - \ln \frac{M^2}{\mu^2} \right) \int d^4 \theta X^{\dag \alpha}_r(-p,\theta) P_T X_\alpha^r(p,\theta),
}
where $M$ is for the masses of chiral superfields running in the internal line. The third contribution (the diagram (c) in \cref{fig:quad}) comes from the vertex which includes the VEV of the adjoint Higgs superfield,
\eqs{
\Gamma_{XX}^{\text{(c)}} = \frac{1}{16\pi^2} A(p^2,M_1^2,M_2^2) & \int d^4 \theta X^{\dag \alpha}_r(-p,\theta) P_T X_\alpha^r(p,\theta) .}
Here, the loop function $A$ is also defined in \cref{eq:loop_func}.

\subsubsection*{Radiative Corrections to Three-Point Vertices}
The one-loop diagrams for the three-point vertex correction  are shown in \cref{fig:ver_corr}. In our momentum assignment, the momentum of the $X$ boson is $q=0$. The one-loop vertex correction induced by the diagram in \cref{fig:ver_corr} (a) is given as 
\eqs{
i\Gamma_{1}^{\rm (v)}(p;M) = & ~ i^3 \int d^4\theta_1 d^4\theta_2 d^4\theta_3 \int \frac{d^D l}{(2\pi)^D} \frac{i}{(l+p)^2}\frac{i}{(l+p)^2} \frac{-i}{2(l^2-M^2)} \\
&\times \frac{1}{16} (\ovl \coD_2^{2} \del_{23} \ola{\coD^{2}_3}) \frac{1}{16} (\ovl \coD_3^{2} \del_{31} \ola{\coD^{2}_1}) \del_{12}
 \Phi(\theta_1) \Phi^\dag(\theta_2) V(\theta_3).
}
By integrating by part and also using the $\coD$ algebra, we always decompose the vertex correction into the effective K\"ahler terms ${\cal K}$ and the auxiliary terms which vanish as $\coD_\alpha \Phi, \ovl \coD_{\dot\alpha}\Phi^\dag = 0$. The effective K\"ahler term induced by the diagram \cref{fig:ver_corr} (a) has the following form:
\eqs{
& i {\cal K}_{1}^{\rm (v)}(p;M) 
 = \frac{1}{2} \int \frac{d^D l}{(2\pi)^D} \frac{1}{[(l+p)^2]^2} \frac{1}{l^2-M^2}(l+2p)^2 \Phi \Phi^\dag V,
}
where we remove the Grassmann valuables in the effective K\"ahler term, for simplicity.

Next we show the effective K\"ahler term described in \cref{fig:ver_corr} (b) and (c). In our momentum assignment, the diagrams both of \cref{fig:ver_corr} (b) and (c) give the same expression, and we find the one-loop vertex correction and the effective K\"ahler term as
\eqs{
i \Gamma_{2}^{\rm (v)}(p;M) = & ~ i^2 \int d^4\theta_1 d^4 \theta_2 \int \frac{d^D l}{(2\pi)^D} \frac{i}{(l+p)^2} \frac{-i}{2(l^2-M^2)} \\
& \times \frac{1}{16}
 \ovl \coD_2^2\del_{21} \ola{\coD^2_1}  \del_{12}  \Phi(\theta_1) \Phi^\dag(\theta_2) V(\theta_2), \\
i {\cal K}_{2}^{\rm (v)}(p;M) 
= & - \frac{1}{2} \int \frac{d^D l}{(2\pi)^D} \frac{1}{(l+p)^2} \frac{1}{l^2-M^2} \Phi \Phi^\dag V.
}

The diagrams (d) and (e) in \cref{fig:ver_corr} include the three-point vertices of vector superfields. After carrying out the superspace integral, the vertex corrections from the diagrams \cref{fig:ver_corr}(d) and (e) are obtained as 
\eqs{
i \Gamma_{3}^{\rm (v)}(p;M) 
& = - 4 \int \frac{d^D l}{(2\pi)^D} \frac{1}{l^2-\mu_{\rm IR}^2} \frac{1}{(l+p)^2} \frac{(l+p)^2 + p^2}{l^2-M^2} \int d^4\theta ~ \Phi^\dag \Phi V, \\
i \Gamma_{4}^{\rm (v)}(p;M) & = 8 \int\frac{d^D l}{(2\pi)^D} \frac{1}{l^2-M^2}\frac{1}{l^2-\mu_{\rm IR}^2} \int d^4\theta ~ \Phi^\dag \Phi V.\\
}
Since they do not include the auxiliary terms, $\Gamma_{n}^{\rm (v)}(p;M)$ ($n=3,4$) is just the K\"ahler term $\int d^4\theta {\cal K}_{n}^{\rm (v)}(p;M)$ ($n=3,4$).

The contribution from a diagram (f) in \cref{fig:ver_corr} is zero as mentioned in the text. 

\subsubsection*{Box-like  Corrections}

Now we show the effective K\"ahler terms from the box-like diagrams presented in \cref{fig:box_corr}. These diagrams include one massless and one massive vector superfields. The correction from the box diagram (\cref{fig:box_corr}(a)) is given as:
\eqs{
i\Gamma_{\rm box} (p;M) 
& \equiv 
p^2 \int \frac{d^D q}{ (2\pi)^D} \frac{1}{q^2-M^2} \frac{1}{(q+p)^2}\frac{1}{q^2-\mu_{\rm IR}^2} \frac{1}{(q-p)^2} \int d^4 \theta \Phi^\dag_1 \Phi_2 \Phi^\dag_3 \Phi_4. \\
}
Here, we do not write the external momenta of external superfields for simplicity since we set them to be the same momentum $p$. As mentioned above, we set the mass of massless vector superfields to be $\mu_{\rm IR}$ as IR regularization. 
$\Gamma_{\rm box} (p;M) $  vanishes at the point with $p^2=0$, as mentioned in the text.

The contribution of the crossing-box diagram (\cref{fig:box_corr}(b)) is given by
\eqs{
i \Gamma_{\rm cross} (p;M) 
& \equiv 
\frac14 \int \frac{d^D q}{ (2\pi)^D} \frac{1}{q^2-M^2} \frac{1}{[(q-p)^2]^2}\frac{1}{q^2-\mu_{\rm IR}^2} 
 \int d^4 \theta \left[ (q-p)^2 \Phi^\dag_1 \Phi_2 \Phi^\dag_3 \Phi_4 \right. \\
& + \frac12  (q-p)_\mu (\ovl\si^\mu \ovl \coD \coD) \left( \Phi^\dag_1 \Phi_4 \right) \Phi_2 \Phi^\dag_3 + \left. \frac1{16} \ovl \coD^2 \coD^2 \left( \Phi^\dag_1 \Phi_4 \right) \Phi^\dag_3 \Phi_2 \right]. \\
}
Here, we define the mnemonic symbol $(\ovl\si^\mu \ovl \coD \coD)  \equiv (\ovl\si^\mu)^{\dot\alpha\alpha} \ovl \coD_{\dot\alpha} \coD_\alpha$.
This correction has the auxiliary terms. The corresponding K\"ahler term is given by removing the auxiliary terms as 
\eqs{
& i {\cal K}_{\rm cross} (p;M) 
 = \frac14 \int \frac{d^D q}{ (2\pi)^D}\frac{(q-2p)^2}{(q^2-M^2)[(q-p)^2]^2 (q^2-\mu_{\rm IR}^2)} \Phi^\dag_1 \Phi_2 \Phi^\dag_3 \Phi_4.
}

Finally, we show the contribution from the triangle diagram in \cref{fig:box_corr}(c). The correction from the triangle diagram is obtained as follows:
\eqs{
i \Gamma_{\rm triangle} (p;M) 
& \equiv
- \frac{1}{4} \int \frac{d^D q}{ (2\pi)^D} \frac{1}{q^2-M^2} \frac{1}{(q+p)^2}\frac{1}{q^2-\mu_{\rm IR}^2} \int d^4 \theta \Phi^\dag_1 \Phi_2 \Phi^\dag_3 \Phi_4. \\
}
Since auxiliary terms are not included in the radiative corrections $\Gamma_{\rm box}$ and $\Gamma_{\rm triangle}$, the corresponding K\"ahler terms are just written by these corrections as $\Gamma_n = \int d^4 \theta ~ {\cal K}_n$ ($n = \text{box, triangle}$).

The diagram in \cref{fig:box_corr}(d) vanishes as mentioned in the text.

\subsubsection*{One-loop Corrections in EFT}
In the last of this appendix, we show the radiative corrections in EFT presented in \cref{fig:sd-eff}. We obtain the one-loop effective vertex functions $\Gamma^{\rm EFT}_1, \Gamma^{\rm EFT}_2$, and $\Gamma^{\rm EFT}_3$ which correspond to the diagram \cref{fig:sd-eff} (b), (c), and (a), respectively, as follows:
\eqs{
 i \Gamma^{\rm EFT}_1 (p;\mu_{\rm IR}) 
& = \frac12 \int \frac{d^D q}{ (2\pi)^D} \frac{1}{[(q+p)^2]^2} \frac{1}{q^2-\mu_{\rm IR}^2} \int d^4 \theta \left[ (q+p)^2 \Phi^\dag_1 \Phi_2 \Phi^\dag_3 \Phi_4 \right.\\
& + \frac12  (q+p)_\mu (\ovl\si^\mu \ovl \coD \coD) \left( \Phi^\dag_1 \Phi_2 \right) \Phi^\dag_3 \Phi_4 + \left. \frac1{16} \ovl \coD^2 \coD^2 \left( \Phi^\dag_1 \Phi_2 \right) \Phi^\dag_3 \Phi_4 \right], \\
i \Gamma^{\rm EFT}_2 (p;\mu_{\rm IR}) 
& = - \frac{1}{2} \int \frac{d^D q}{ (2\pi)^D} \frac{1}{q^2-\mu_{\rm IR}^2} \frac{1}{(q+p)^2}\int d^4 \theta \Phi^\dag_1 \Phi_2 \Phi^\dag_3 \Phi_4, \\
 i \Gamma^{\rm EFT}_3(p;\mu_{\rm IR}) 
& = \frac12 \int \frac{d^D q}{ (2\pi)^D} \frac{1}{[(q+p)^2]^2} \frac{1}{q^2-\mu_{\rm IR}^2} (2p)^2 \int d^4 \theta \Phi^\dag_1 \Phi_2 \Phi^\dag_3 \Phi_4.
}
The momentum assignment is the same as in calculation of the box-like diagrams.
The corresponding K\"ahler terms are given by removing the auxiliary terms as 
\eqs{
i{\cal K}^{\rm EFT}_1(p;\mu_{\rm IR}) & = \frac12 \int \frac{d^D q}{ (2\pi)^D}\frac{(q+2p)^2}{[(q+p)^2]^2 (q^2-\mu_{\rm IR}^2)} \Phi^\dag_1 \Phi_2 \Phi^\dag_3 \Phi_4, \\
i {\cal K}^{\rm EFT}_2(p;\mu_{\rm IR}) & = - \frac{1}{2} \int \frac{d^D q}{ (2\pi)^D} \frac{1}{q^2-\mu_{\rm IR}^2} \frac{1}{(q+p)^2} \Phi^\dag_1 \Phi_2 \Phi^\dag_3 \Phi_4, \\
 i {\cal K}^{\rm EFT}_3(p;\mu_{\rm IR}) 
& = 2p^2 \int \frac{d^D q}{ (2\pi)^D} \frac{1}{[(q+p)^2]^2} \frac{1}{q^2-\mu_{\rm IR}^2}  \Phi^\dag_1 \Phi_2 \Phi^\dag_3 \Phi_4. \label{eq:kahler_EFT}
}
Here, we skip over the ways in which we obtain the effective vertex functions from the diagrams since these structures are similar as mentioned above. $i{\cal K}^{\rm EFT}_3(p;\mu_{\rm IR})$ vanishes when $p^2= 0$ is set, as mentioned in the text.

\section{Renormalization Group Equations\label{app:RGE}}

\subsubsection*{Gauge Couplings and Yukawa Couplings}

In our analysis, we have used the RGEs at the two-loop level. The RGEs for the gauge coupling constants are as follows \cite{Machacek:1983tz,Martin:1993zk}:
\eqs{
\diff{1}{g_i}{\ln \mu} = \frac{g_i}{16\pi^2} \left[ b_i g_i^2 + \frac{1}{16\pi^2} \left( \sum_j b_{ij} g_i^2 g_j^2 - \sum_{j=U,D,E} a_{ij} g_i^2 \tr[\bold Y_j \bold Y_j^\dag ] \right) \right ].
}
Here, $\bold Y_U =\bold U, \bold Y_D = \bold D$, and $\bold Y_E={\bold E}$ are the Yukawa coupling matrices. The coefficients in the SM are given as:
\eqs{
b_{ij} = \left(
\begin{array}{ccc}
\frac{199}{50} & \frac{27}{10}  & \frac{44}{5}  \vspace{+.5em}\\
\frac{9}{10} &  \frac{35}{6} & 12  \vspace{+.5em}\\
\frac{11}{10} & \frac92  & -26  
\end{array}
\right), ~~~
b_i = \left(\frac{41}{10},-\frac{19}{6},-7\right), ~~~
a_{ij} = \left(
\begin{array}{ccc}
 \frac{17}{10} & \frac12  & \frac32  \vspace{+.5em}\\
 \frac32 &  \frac32 & \frac12  \vspace{+.5em}\\
 2 & 2  & 0
\end{array}
\right).\label{eq:beta}
}
The one-loop RGEs for the Yukawa coupling matrices are given as\footnote{In our calculation, we need the RGEs for the gauge couplings at the two-loop level. It is sufficient to take into account the RGEs for the Yukawa couplings at the one-loop level since the Yukawa couplings appear in the two-loop-level RGEs for the gauge couplings.}.
\eqs{
\diff{1}{\bold U}{\ln \mu} & = \frac{1}{16\pi^2} \left[ - \sum_i c^{\rm SM}_i g_i^2 + \frac{3}{2} \bold U \bold U^\dag - \frac{3}{2} \bold  D \bold  D^\dag + Y_2(S) \right] \bold U, \\
\diff{1}{\bold D}{\ln \mu} & = \frac{1}{16\pi^2} \left[ - \sum_i c'^{\rm SM}_i g_i^2 + \frac{3}{2} \bold D \bold D^\dag - \frac{3}{2} \bold  U \bold  U^\dag + Y_2(S) \right] \bold D, \\
\diff{1}{\bold E}{\ln \mu} & = \frac{1}{16\pi^2} \left[ - \sum_i c''^{\rm SM}_i g_i^2 + \frac{3}{2} \bold E \bold E^\dag + Y_2(S) \right] \bold E, \\
}
where
\eqs{
c^{\rm SM}_i = \left( \frac{17}{20} , \frac{9}{4} , 8 \right), ~~~
c'^{\rm SM}_i = \left( \frac{1}{4} , \frac{9}{4} , 8 \right), ~~~
c''^{\rm SM}_i = \left( \frac{9}{4} , \frac{9}{4} , 0 \right),
}
and 
\eqs{
Y_2(S) = \tr \left[ 3 \bold U \bold U^\dag + 3 \bold D \bold D^\dag + \bold E \bold E^\dag \right].
}

The coefficients of the RGEs for the gauge coupling constants in the MSSM are obtained as
\begin{equation}
\begin{split}
b_{ij} = \left(
\begin{array}{ccc}
\frac{199}{25} & \frac{27}{5}  & \frac{88}{5}  \vspace{+.5em}\\
\frac95 &  25 & 24 \vspace{+.5em}\\
\frac{11}{5} & 9  & 14
\end{array}
\right), ~~~
b_i = \left(\frac{33}{5},1,-3\right), ~~~
a_{ij} = \left(
\begin{array}{ccc}
\frac{26}{5} & \frac{14}{5}  & \frac{18}{5}  \vspace{+.5em}\\
 6 &  6 & 2  \vspace{+.5em}\\
 4 & 4  & 0
\end{array}
\right).
\end{split}
\end{equation}
The one-loop RGEs for the Yukawa matrices in the MSSM  are given as 
\eqs{
\diff{1}{\widetilde{\bold U}}{\ln \mu} & = \frac{1}{16\pi^2} \left[ - \sum_i c^{\rm MSSM}_i g_i^2 + 3 \widetilde{\bold U} \widetilde{\bold U}^\dag + \widetilde{\bold D} \widetilde{\bold D}^\dag + \tr(3 \widetilde{\bold U} \widetilde{\bold U}^\dag) \right] \widetilde{\bold U}, \\
\diff{1}{\widetilde{\bold D}}{\ln \mu} & = \frac{1}{16\pi^2} \left[ - \sum_i c'^{\rm MSSM}_i g_i^2 + 3 \widetilde{\bold D} \widetilde{\bold D}^\dag + \widetilde{\bold U} \widetilde{\bold U}^\dag + \tr(3 \widetilde{\bold D} \widetilde{\bold D}^\dag + \widetilde{\bold E} \widetilde{\bold E}^\dag) \right] \widetilde{\bold D}, \\
\diff{1}{\widetilde{\bold E}}{\ln \mu} & = \frac{1}{16\pi^2} \left[ - \sum_i c''^{\rm MSSM}_i g_i^2 +3\widetilde{\bold E} \widetilde{\bold E}^\dag + \tr(3 \widetilde{\bold D} \widetilde{\bold D}^\dag + \widetilde{\bold E} \widetilde{\bold E}^\dag) \right] \widetilde{\bold E}, \\
}
where 
\eqs{
c^{\rm MSSM}_i = \left( \frac{13}{15} , 3 , \frac{16}{3} \right), ~~~
c'^{\rm MSSM}_i = \left( \frac{7}{15} , 3 , \frac{16}{3} \right), ~~~
c''^{\rm MSSM}_i = \left( \frac{9}{5} , 3 , 0 \right).
}
The boundary conditions for the Yukawa coupling constants at the SUSY breaking scale ($M_S$) are 
\eqs{
\widetilde U(M_S) = \frac{1}{\sin \beta} U(M_S), ~~~~
\widetilde Y_j(M_S) = \frac{1}{\cos \beta}Y_j(M_S) ~~ (j = D,E).
}
where $\tan\beta$ is the ratio of vacuum expectation values in the MSSM.

When the vector-like matters are introduced in the MSSM, the RGEs for the gauge coupling constants are modified as
\eqs{
b_i \to b_i + \del b_i, ~~~~~~
b_{ij} \to b_{ij} + \del b_{ij},
}
where $b_i$ and $b_{ij}$ are the coefficients of the one-loop and two-loop RGEs in the MSSM, respectively.
$\del b_i$ and $\del b_{ij}$ are given by \cite{Ghilencea:1997mu}:
\eqs{
\del b_i & = \left( n_{\bold{5}}+3n_{\bold{10}}, n_{\bold{5}}+3n_{\bold{10}}, n_{\bold{5}}+3n_{\bold{10}} \right), \\
\del b_{ij} & = \left(
\begin{array}{ccc}
\frac{7}{15} n_{\bold{5}} + \frac{23}{5} n_{\bold{10}} & \frac95 n_{\bold{5}} + \frac35 n_{\bold{10}}  & \frac{32}{15} n_{\bold{5}} + \frac{48}{5} n_{\bold{10}} \vspace{+.5em}\\
\frac35 n_{\bold{5}} + \frac15 n_{\bold{10}} &  7 n_{\bold{5}} + 21n_{\bold{10}} & 16 n_{\bold{10}} \vspace{+.5em}\\
\frac{4}{15}n_{\bold{5}} + \frac65n_{\bold{10}} & 6 n_{\bold{10}}  & \frac{34}{3} n_{\bold{5}} + 34 n_{\bold{10}}
\end{array}
\right),
}
where $n_{\bold{5}}$ and $n_{\bold{10}}$ denote the number of $\bold{5}+\ovl{\bold{5}}$ and $\bold{10}+\ovl{\bold{10}}$ vector-like matter superfields, respectively.

\subsubsection*{Wilson Coefficients of $D=6$ Baryon-Number Violating Operators}
In Ref.~\cite{Hisano:2013ege}, they have derived the two-loop RGEs for the Wilson coefficients of the following dimension-six baryon-number violating operators in the SUSY invariant theories,
\eqs{
{\cal L}_{D=6} = \sum_{I=1}^2 {\cal C}^{(I)} {\cal O}^{(I)},
}
where 
\eqs{
{\cal O}^{(1)} =  & ~ \ep_{\alpha\beta\gamma} \ep_{rs} \int d^4 \theta  U^{C\dag \alpha} D^{C\dag\beta} e^{-\frac23 g_Y B} (e^{2g_3 G} Q^r)^{\gamma} L^s, \\
{\cal O}^{(2)} =& ~ \ep_{\alpha\beta\gamma} \ep_{rs} \int d^4 \theta  E^{C\dag} (e^{-2g_3 G} U^{C\dag})^{\alpha} e^{\frac23 g_Y B}  Q^{r \beta}Q^{s\gamma}.
}
The RGEs for the Wilson coefficients are given as 
\eqs{
\mu \diff{1}{{\cal C}^{(I)}}{\mu} = \frac{{\cal C}^{(I)}}{16\pi^2}  \left[ \sum_i \alpha^{(I)}_i g_i^2 + \frac{1}{16\pi^2}  \sum_{i,j} \alpha^{(I)}_{ij} g_i^2 g_j^2 \right ], 
}
where $i=1,\cdots3$, and the coefficients are given as
\eqs{
\alpha_i^{(1)} = \left(-\frac{11}{15}, -3, - \frac83 \right), ~~~~~~
\alpha_i^{(2)} = \left(-\frac{23}{15}, -3, - \frac83 \right),
}
\eqs{
\alpha^{(1)}_{ij} = \left(
\begin{array}{ccc}
\frac{113}{150} + b_1 & \frac35  & \frac{34}{15}  \vspace{+.5em}\\
\frac35 &  \frac92+3b_2 & 6 \vspace{+.5em} \\
\frac{34}{15} & 6  & \frac{64}{3} + 8 b_3
\end{array}
\right), ~~~~~~
\alpha^{(2)}_{ij} = \left(
\begin{array}{ccc}
\frac{91}{50} +\frac95 b_1 & \frac15  & \frac{38}{15}  \vspace{+.5em}\\
\frac15 &  \frac92+3b_2 & 10 \vspace{+.5em}\\
\frac{38}{15} & 10  & \frac{64}{3} + 8 b_3
\end{array}
\right).
}
Here, $b_i~(i=1$-$3)$ is given in \cref{eq:beta}.

\newpage
\bibliography{ref}
\end{document}